\documentclass[11pt,a4paper]{article}

\usepackage{amsmath}
\usepackage{amsfonts}
\usepackage{color}
\usepackage{cite}
\usepackage{graphicx}
\usepackage{float}
\usepackage{subfigure}

\title{\textbf{Families of Q-balls in a deformed $O(4)$ linear sigma model}}

\author{A. Alonso-Izquierdo$^{(a,b)}$ and C. Garz\'on S\'anchez$^{(a)}$
\\ [1ex]
$^{(a)}$ Departamento de Matematica Aplicada, University of Salamanca, \\ Casas del Parque 2, 37008 - Salamanca, Spain \\ [1ex]
$^{(b)}$  IUFFyM, University of Salamanca, \\ Plaza de la Merced 1, 37008 - Salamanca, Spain \\ [1ex]}

\date{\today}

\begin{document}

\maketitle

\begin{abstract}
In this paper the existence of analytical solutions describing $Q$-balls in a family of deformed $O(4)$ sigma models in (1+1) dimensions has been investigated. These models involve two complex scalar fields whose coupling breaks the $O(4)$ symmetry group to $U(1)\times U(1)$. It has been shown that there are two types of single $Q$-balls rotating around each of the components of the internal space and a one-parameter family of composite $Q$-balls. These composite solutions consist of two single $Q$-balls (separated by a distance determined by the family parameter) spinning around each complex field with the same internal rotation frequency. 
\end{abstract}

\section{Introduction}

$Q$-balls are time-dependent non-topological solutions defined in nonlinear field theories characterized by the presence of a global $U(1)$ symmetry \cite{Shnir2018}. This symmetry gives place to a conserved Noether charge, which is associated to an angular motion with angular velocity $\omega$ in the internal space. The time dependence of these solutions allows them to avoid the severe restrictions of Derrick's theorem \cite{Derrick1964} whereas the internal rotation stabilizes the $Q$-balls, which in other case would decay to the vacuum solution \cite{Lee1992, Dine2003,Tsumagari2008}. Other possibilities for stabilizing non-topological defects in (1+1)-dimensions by introducing a potential barrier or a non-trivial target space have been described in \cite{Alonso2007,Alonso2020,Alonso2022}. In 1976 Friedberg, Lee and Sirlin studied the presence of this type of solutions in a theoretical model involving a complex scalar field coupled with a real scalar field in the seminal paper \cite{Friedberg1976}. The authors describe the $Q$-balls arising in this model and thoroughly analyze the stability of these solutions versus small fluctuations that maintain the conserved Noether charge constant. After this pioneering work, the existence of $Q$-balls and its properties have been studied in different contexts, for example, in complex scalar field theories \cite{Coleman1985, Coleman1986,Kusenko1997, Nugaev2014}, in Abelian gauge theories \cite{Lee1989, Anagnostopolos2001, Li2001,Panin2019}, in non-Abelian theories \cite{Safian1988, Safian1988b, Axenides1998}, in models which include fermionic interactions \cite{Levi2002,Friedbert1977, Friedbert1977b, Cohen1986, Anagnostopolos2001}, in models with presence of gravity \cite{Lee1987, Lynn1989}, etc. This interest is explained by the fact that $Q$-balls are thought to play a relevant role in some important physical phenomena. For example, in 1998 Kusenko and Shaposhnikov \cite{Kusenko1998} showed that $Q$-balls can be produced in the early universe in supersymmetric extensions of the standard model in such a way that they can contribute to dark matter by means of the Affleck-Dine mechanism. The relevance of this fact is that it is conjectured that this mechanism could explain baryogenesis during the primordial expansion, after cosmic inflation. It has also been proposed that $Q$-balls present in models with gravity mediated supersymmetry breaking are long-lived, allowing in principle, for these $Q$-balls to be the source of both the baryons and the lightest supersymmetric particle dark matter particle \cite{Enqvist2003}.

In general, models involving $Q$-balls are so complicated that it is not possible to obtain analytical expressions for these non-topological solitons. For this reason an interesting research direction in the study of $Q$-balls is that of identifying models where these solutions can be analytically calculated. This knowledge opens the possibility of analyzing their properties further and in more detail. This scenario has been explored in theories with one complex scalar field in (1+1)-dimensions in recent works \cite{Bazeia2016, Bazeia2016b, Bazeia2017, Bazeia2019}. Here, the authors focus mainly on attaining analytical solutions with different features. For example, models involving compact $Q$-balls  are constructed in \cite{Bazeia2016b}. Another relevant topic deals with the study of excited states of $Q$-balls. In \cite{Shiromizy1998, Shiromizu1999} the authors investigate this class of configurations by performing stationary perturbations on spherical $Q$-balls and describe the implications to Cosmology derived from the magnetic fields generated by these excited solutions. Radial excitations are examined in \cite{Volkov2002} in the case of models with one complex scalar field. Additionally, vibrational modes of these solutions with a near-critical charge or in theories with flat potential are analyzed in \cite{Kovtun2018}. On the other hand, the interaction between these non-topological solitons leads to a very rich variety of behaviors. For example, processes such as charge transfer and fission has been identified in one, two and three space dimensions \cite{Battye2000}. The scattering between $Q$-balls in (1+1)-dimensions has also been investigated in \cite{Bowcock2009}. Curiously, it was found that attractive or repulsive forces arise depending upon the relative phase of the colliding $Q$-balls. 

In this paper we shall investigate the existence of analytical non-topological solitons in a family of two-component complex scalar field theories in (1+1) dimensions with a global $U(1)\times U(1)$ symmetry. The coupling between the two fields breaks down a $O(4)$ symmetry in such a way that the model parameter can be understood as a measure of the deformation of the model with respect to the rotationally invariant theory. The effect of introducing two complex fields have been previously explored by Loginov and Gauzshtein \cite{Loginov2018,Loginov2018b, Loginov2019} although in a different framework. For example, in \cite{Loginov2018,Loginov2018b} the authors deal with a (2+1)-dimensional gauge model describing two complex scalar fields that interact through a common Abelian gauge field. In this framework composite solutions arise consisting of a vortex and a $Q$-ball. A similar model immersed in a (1+1)-dimensional space-time is addressed in \cite{Loginov2019}. It is shown that the model has again composite solutions consisting now of two $Q$-balls with opposite electric charges. Note that the Friedberg-Lee-Sirlin model \cite{Friedberg1976} involves the coupling between a real and a complex field. It has been shown the existence of hairy $Q$-balls in the limiting case of the vanishing potential in the previously mentioned model \cite{Loiko2018}. Composite $Q$-balls involving different geometries have been studied in \cite{Shnir2011}. 

Obviously, the study of soliton solutions in these coupled theories is much more complex than the case where only one field is present. Despite this fact, $Q$-ball solutions of the models introduced in this paper are analytically identified. As previously mentioned the model involves the presence of two complex fields and exhibits a $U(1)\times U(1)$ symmetry, which gives place to the existence of two conserved Noether charges. In this scenario a general $Q$-ball is described by the profiles of these two fields, which in principle could rotate with different internal frequencies. In particular, it will be shown that for certain ranges of the internal rotation frequencies there are two types of single $Q$-balls rotating around one of the complex fields while the other field vanishes. These single energy lump solutions are stable. In addition, there exists a one-parameter family of composite $Q$-balls when the two previously mentioned internal rotation frequencies are synchronized. These solutions consist of two single $Q$-balls and are formed by two energy lumps separated by a distance determined by the family parameter. This scheme suggests that, for configurations where the internal rotation frequencies are different, forces between the different constituents of the solutions arise making the non-topological solutions depend non-trivially on time. The study of the stability of these composite solutions is very complicated and some of the arguments introduced in \cite{Friedberg1976} must be altered when applied to these models. For example, the existence of two negative eigenvalues in the spectrum of the second order small fluctuation operator is not a sufficient condition for proving the instability of the $Q$-balls. This argument is now valid if three of these eigenstates are considered. Numerical analysis seems to indicate that these composite solutions are long-lived but unstable states. The analysis of these instability channels could bring insight into the forces between the two single $Q$-balls when they approach each other.

The organization of this paper is as follows: the family of deformed $O(4)$ linear sigma models addressed in this work and its properties is introduced in Section 2. The previously mentioned single $Q$-balls and the family of composite $Q$ balls are analytically identified and described in Section 3. Sum rules between the energies and the conserved charges of these solutions are also discussed. Section 4 is dedicated to investigate the linear stability of these non-topological solitons. Finally, the conclusions of this work are summarized in Section 5.

\section{The model} \label{Model}

\begin{figure}[h]
	\centerline{\includegraphics[height=3.cm]{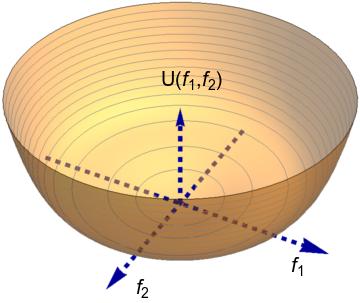} \hspace{0.5cm}\includegraphics[height=3.cm]{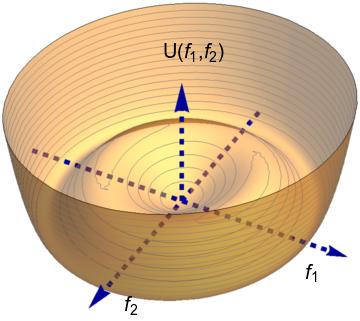} \hspace{0.5cm} \includegraphics[height=3.cm]{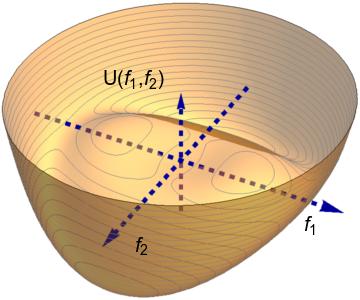} \hspace{0.5cm} \includegraphics[height=3.cm]{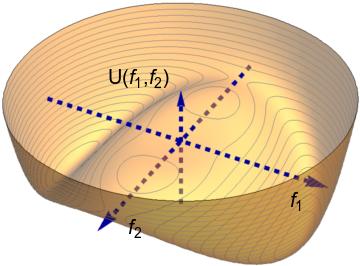}}
	\caption{\small Graphics of the effective potential $\overline{U}(f_1,f_2)$ for the parameter values $\sigma=0.25$, $a=1.75$, $b=2.0$ and several values of the internal rotation frequencies: (a) $\omega_1=\omega_2=0$, (b) $\omega_1=\omega_2=1.29$, (c) $\omega_1=1.29$, $\omega_2=1.75$ and (d) $\omega_1=1.95$, $\omega_2=1.3$.} \label{fig:potential}
\end{figure}

We shall deal with a two-component complex scalar field theory immersed in a Minkowski spacetime characterized by the action functional
\begin{equation}
S=\int d^2 x \Big[ \frac{1}{2} \partial_\mu \overline{\phi} \, \partial^\mu \phi +  \frac{1}{2} \partial_\mu \overline{\psi} \, \partial^\mu \psi - U(|\phi|,|\psi|) \Big] \hspace{0.4cm}, \label{action}
\end{equation}
where $\phi=\phi_1+i\phi_2$ and $\psi=\psi_1+i\psi_2$ are dimensionless complex scalar fields, that is, $\phi,\psi \in {\rm Maps} (\mathbb{R}^{1,1}, \mathbb{C} )$ and $\overline{\phi}$ stands for complex conjugate of $\phi$. The Minkowski metric $g_{\mu\nu}$ is chosen as $g_{00}=-g_{11}=1$ and $g_{12}=g_{21}=0$. In order to alleviate the notation we introduce the two-component complex scalar field $\Phi=( \phi, \psi )$, which let us define
\begin{equation}
|\Phi|^2 = |\phi|^2 + |\psi|^2 = |\phi_1|^2 + |\phi_2|^2 + |\psi_1|^2+ |\psi_2|^2 \hspace{0.4cm}. \label{modulo}
\end{equation}
With this notation, the potential term $U(|\phi|,|\psi|)$ which will be investigated in this paper can be written as
\begin{equation}
U(|\phi|,|\psi|; \sigma,a,b) = \frac{1}{2} \Big( |\Phi|^6 - a^2 |\Phi|^4 + b^2 |\Phi|^2 + 2 \sigma^2 |\psi|^2 |\Phi|^2 +  \sigma^2 (\sigma^2-a^2) |\psi|^2 \Big)  \label{potential}
\end{equation}
with $a,b,\sigma\in \mathbb{R}$. The relation (\ref{potential}) is a sixth-degree algebraic expression in the modulus of the complex fields $\phi$ and $\psi$. It is a deformation of the $O(4)$-invariant $|\Phi|^6$-model, where the parameter $\sigma$ measures the asymmetry with respect to the rotationally invariant situation. This can be checked by noting that for $\sigma=0$
\begin{equation}
U(|\phi|,|\psi|; 0,a,b) = \frac{1}{2} ( |\Phi|^6 - a^2 |\Phi|^4 + b^2 |\Phi|^2 ) \hspace{0.4cm}. \label{potential0}
\end{equation}
On the other hand, $a$ and $b$ are the usual parameters that allow to change the profile of the $|\Phi|^6$-potential \cite{Shnir2018}. The potential (\ref{potential}) has a critical point at $(\phi,\psi)=(0,0)$, where the potential vanishes, $U(0,0)=0$. The Hessian matrix of (\ref{potential}) evaluated at this point is given by
\[
{\cal H}[0,0] = \left. \left( \begin{array}{cc} \frac{\partial^2 U}{\partial |\phi|^2} & \frac{\partial^2 U}{\partial |\phi| \partial |\psi|} \\[0.2cm] \frac{\partial^2 U}{\partial |\phi| \partial |\psi|} & \frac{\partial^2 U}{\partial |\psi|^2} \end{array} \right) \right|_{(0,0)} =  \left( \begin{array}{cc} b^2  & 0 \\ 0 & b^2+ \sigma^2 (\sigma^2 - a^2)   \end{array} \right) \hspace{0.4cm},
\]
which means that $(\phi,\psi)=(0,0)$ is a minimum point only if the condition
\begin{equation}
b^2 > \sigma^2(a^2 -\sigma^2) \label{condition01}
\end{equation}
holds. We are interested in searching for $Q$-ball type solutions, so we shall restrict our study to this regime, which guarantees the linear stability of the vacuum solution $\Phi=0$. As previously mentioned, the $O(4)$-symmetry associated to (\ref{potential0}) is broken in our model for $\sigma\neq 0$, see (\ref{potential}). However, it is not completely broken and two $U(1)$-symmetries involving each of the complex fields are still preserved. The model is invariant with respect to the transformations $\phi \rightarrow e^{i\alpha} \phi$ and $\psi \rightarrow e^{i\beta} \psi$, which leads to the conserved Noether charges
\begin{equation}
Q_1= \frac{1}{2i} \int \left( \, \overline{\phi} \,\partial_t\phi - \phi \, \partial_t \overline{\phi}  \, \right) dx \hspace{1cm},\hspace{1cm} Q_2= \frac{1}{2i} \int \left( \, \overline{\psi} \,\partial_t\psi - \psi \, \partial_t \overline{\psi} \, \right) dx \hspace{0.4cm}. \label{charges0}
\end{equation}
The field equations obtained from the Euler-Lagrange equations associated to the action funcional (\ref{action}) can be written as
\begin{equation}
\frac{\partial^2\phi}{\partial t^2} - \frac{\partial^2 \phi}{\partial x^2} + \frac{\phi}{|\phi|} \frac{\partial U(|\phi|,|\psi|)}{\partial |\phi|} = 0 \hspace{0.4cm},\hspace{0.4cm}
\frac{\partial^2\psi}{\partial t^2} - \frac{\partial^2 \psi}{\partial x^2} + \frac{\psi}{|\psi|} \frac{\partial U(|\phi|,|\psi|)}{\partial |\psi|} = 0 \hspace{0.4cm} .
\label{pde01}
\end{equation}
In this framework, $Q$-balls are time-dependent regular localized solutions of (\ref{pde01}) which rotates with constant frequency in the internal space of each complex field. They can be studied by substituting the ansatz
\begin{equation}
\phi(x,t)= f_1(x) \, e^{i\omega_1 t} \hspace{1cm},\hspace{1cm} \psi(x,t)= f_2(x) \, e^{i\omega_2 t} \label{ansatz}
\end{equation}
into the field equations (\ref{pde01}). This leads to the ordinary differential equations
\begin{equation}
 \frac{\partial^2 f_1}{\partial x^2} = \frac{\partial U(f_1,f_2)}{\partial f_1} - \omega_1^2 f_1
 \hspace{0.6cm},\hspace{0.6cm}
 \frac{\partial^2 f_2}{\partial x^2} = \frac{\partial U(f_1,f_2)}{\partial f_2} - \omega_2^2 f_2  \label{edo01}
\end{equation}
for the real functions $f_1(x)$ and $f_2(x)$. The quantities $\omega_1$ and $\omega_2$ in (\ref{ansatz}) are respectively the internal rotation frequencies for the $\phi$ and $\psi$ fields. Without loss of generality we can consider that $\omega_1$ and $\omega_2$ are non-negative. The potential term $U$ in (\ref{edo01}) reads now
\begin{equation}
U(f_1,f_2; \sigma, a ,b) =   \frac{1}{2} \Big( |F|^6 - a^2 |F|^4 + b^2 |F|^2 + 2 \,\sigma^2 f_2^2 |F|^2 +  \sigma^2 (\sigma^2-a^2) f_2^2 \Big) \label{potential03}
\end{equation}
where $F=(f_1,f_2)$ and $|F|^2 = f_1^2 + f_2^2$. The integral over the space coordinate of the energy density
\begin{equation}
{\cal E}[f_1,f_2] =  \frac{1}{2} \Big(\frac{\partial f_1}{d x} \Big)^2 +  \frac{1}{2} \Big(\frac{\partial f_2}{d x} \Big)^2 + \frac{1}{2} \omega_1^2 f_1 ^2 + \frac{1}{2} \omega_2^2 f_2 ^2 + U(f_1,f_2; \sigma, a ,b)  \label{energydensity}
\end{equation}
provides us with the total energy, i.e., $E[f_1,f_2] = \int_{-\infty}^\infty {\cal E}[f_1,f_2] \, dx$. Finally, the conserved Noether charges (\ref{charges0}) become
\begin{equation}
Q_1= \omega_1  \int_{-\infty}^\infty (f_1(x))^2  dx \hspace{0.6cm},\hspace{0.6cm} Q_2= \omega_2  \int_{-\infty}^\infty (f_2(x))^2  dx \label{charges}
\end{equation}
in this case. $Q$-balls are finite energy solutions, which implies that the following asymptotic conditions
\begin{equation}
\lim_{x\rightarrow \pm \infty} f_i = \lim_{x\rightarrow \pm \infty} \frac{d f_i}{dx} = 0 \hspace{0.5cm} \mbox{with} \hspace{0.5cm} i=1,2,
\label{asymptotics}
\end{equation}
must hold. It is also clear from (\ref{energydensity}) that the problem involves the effective potential
\begin{equation}
\overline{U}(f_1,f_2; \sigma, a ,b,\omega_1,\omega_2) =  U(f_1,f_2; \sigma, a ,b) -\frac{1}{2} \, \omega_1^2 \, f_1^2 - \frac{1}{2} \, \omega_2^2 \, f_2^2 \label{potential06}
\end{equation}
in such a way that the equations (\ref{edo01}) can be written in the more compact form
\begin{equation}
	\frac{\partial^2 f_1}{\partial x^2} =   \frac{\partial \overline{U}(f_1,f_2)}{\partial f_1}  \hspace{0.6cm},\hspace{0.6cm}
	\frac{\partial^2 f_2}{\partial x^2} = \frac{\partial \overline{U}(f_1,f_2)}{\partial f_2} \hspace{0.4cm} . \label{edo2}
\end{equation}
The effective potential (\ref{potential06}) depends on the internal rotation frequencies. In Figure \ref{fig:potential} the potential $\overline{U}(f_1,f_2)$ has been depicted for several values of $\omega_1$ and $\omega_2$ with fixed values of the rest of the parameters. Note that for $\omega_1=\omega_2=0$ the original potential $U(f_1,f_2)$ has an absolute minimum located at the origin of the internal space but in other cases this point becomes only a local minimum for the effective potential $\overline{U}(f_1,f_2)$. This is a necessary condition for the existence of $Q$-balls.

Solving the system (\ref{edo2}) together with the conditions (\ref{asymptotics}) is tantamount to finding solutions asymptotically beginning and ending at the origin for Newton equations in which $x$ plays the role of time, the particle position is determined by $(f_1,f_2)$ and the potential energy of the particle is $V (f_1,f_2) = - U (f_1,f_2)$. We shall exploit this mechanical analogy in the next Section by using the Hamilton-Jacobi theory in this context. Note that the equations (\ref{edo2}) can be derived from the static effective functional
\begin{equation}
\overline{E}[f_1,f_2] = \int dx \left[ \frac{1}{2} \Big(\frac{\partial f_1}{d x} \Big)^2 +  \frac{1}{2} \Big(\frac{\partial f_2}{d x} \Big)^2 + \overline{U}(f_1,f_2; \sigma, a ,b) \right] \hspace{0.4cm}.  \label{reducedLagran}
\end{equation}
Another consequence of the previous interpretation is that
\begin{equation}
I_1= \frac{1}{2} \Big( \frac{df_1}{dx} \Big)^2 + \frac{1}{2} \Big( \frac{df_2}{dx} \Big)^2 +\frac{1}{2} \, \omega_1^2 \, f_1^2 + \frac{1}{2} \, \omega_2^2 \, f_2^2- U(f_1,f_2; \sigma, a ,b) \label{firstintegral}
\end{equation}
is a first integral of the system (\ref{edo2}). The asymptotic conditions (\ref{asymptotics}) impose that $Q$-balls are characterized by the relation $I_1=0$.

\section{Families of $Q$-balls}

\label{sectionSol}

In this Section we shall investigate the existence of $Q$-ball solutions in the model (\ref{potential}) introduced in Section 1. It is clear that if one of the complex fields vanishes then the reduced potential becomes the usual sixth-order polynomial studied in the literature, see \cite{Shnir2018}. Consequently, the presence of two types of solutions is expected: $Q$-balls whose $\psi$-component is zero and $Q$-balls which comply with the condition $\phi=0$. They will be denoted respectively as $\overline{{\cal B}}_1(x)$ and $\widehat{{\cal B}}_1(x)$. The subscript in the previous notation is used to indicate that they are single $Q$-balls involving only one energy lump, as will be seen later. As it is well known only internal rotational frequencies $\omega_1$ and $\omega_2$ belonging to certain intervals can lead to these solutions. Assuming that the first type of $Q$-balls arises for $\omega_1\in (\omega_1^{-}, \omega_1^{+})$ and the second one for $\omega_2\in (\omega_2^{-}, \omega_2^{+})$, it will be proved that for our model the lowest value of these frequencies coincides, $\omega_1^{-}=\omega_2^{-}$, whereas the highest allowed frequencies verify $\omega_2^{+} \leq \omega_1^{+}$.

However, the main novelty of this model is the existence of $Q$-balls where the two complex fields are non-null. It will be shown that when the two internal rotational frequencies $\omega_1$ and $\omega_2$ are synchronized, that is, $\omega_1=\omega_2=\omega$, a one-parameter family of $Q$-balls arises. Indeed, they can be analytically identified for every value of $\omega$. Every member of this family can be interpreted as a non-linear combination of one $\overline{{\cal B}}_1(x)$-type $Q$-ball and one $\widehat{{\cal B}}_1(x)$-type $Q$-ball, which are separated by a distance determined by the family parameter $\gamma_1$. We will denote these solutions as ${\cal B}_2(x;\gamma_1)$, where the subscript 2 in this notation emphasizes the fact that they are composite $Q$-balls displaying two distinct energy lumps. It will be proved that all the members of the family have both the same total energy $E$ and the same sum of the Noether charges $Q=Q_1+Q_2$.

\subsection{Single $Q$-balls}

As previously mentioned, there are two types of single $Q$-balls:

\begin{itemize}
\item \textit{$\overline{{\cal B}}_1(x)$-type $Q$-balls:} A $|\phi|^6$-model is embedded in our model when $\psi=0$. In this case the effective potential (\ref{potential06}) reads
\begin{equation}
\overline{U}(f_1, 0; \sigma, a ,b,\omega_1,\omega_2) =\frac{1}{2} \Big( f_1^6 - a^2 f_1^4 + b^2 f_1^2\Big) -\frac{1}{2} \, \omega_1^2 \, f_1^2 \hspace{0.5cm} . \label{pot01}
\end{equation}
The values of the rotational frequency $\omega_1$ are restricted by the following conditions: (1) The effective potential (\ref{pot01}) must have a minimum at $f_1=0$ and (2) the effective potential (\ref{pot01}) must have at least one nontrivial real root. Taking into account that
\[
\frac{\partial^2 \overline{U}(f_1, 0)}{\partial f_1^2} \Big|_{f_1=0}= b^2-\omega_1^2
\]
and that the roots of (\ref{pot01}) are determined by
\begin{equation}
\widetilde{f}_1 = \pm \frac{1}{\sqrt{2}} \Big( a^2 \pm \sqrt{a^4 - 4 (b^2 - \omega_1^2)} \,\, \Big)^\frac{1}{2} \label{roots01}
\end{equation}
the previous conditions restrict the values of $\omega_1$ as follows:
\begin{equation}
 b^2 - \textstyle{\frac{1}{4}} \, a^4 \leq \, \omega_1^2 \, \leq b^2  \hspace{0.4cm} . \label{rango01}
\end{equation}
For this type of solutions the second equation of the system (\ref{edo2}) (or equivalently (\ref{edo01})) is automatically satisfied. Solving the first equation leads to the solution
\begin{equation}
\overline{{\cal B}}_1(x;\omega_1) =(f_1(x),f_2(x)) = \left(\frac{\sqrt{2(b^2-\omega_1^2)}}{\sqrt{a^2 + \sqrt{a^4- 4(b^2-\omega_1^2)} \cosh (2\sqrt{b^2-\omega_1^2} \, \overline{x}) }}\, , \,0 \right) \label{qball01}
\end{equation}
where $\overline{x}=x-x_0$ and $x_0$ can be interpreted as the $Q$-ball center. The Noether charges for this solution are given by
\begin{equation}
Q_1 [ \overline{{\cal B}}_1(x) ] = \omega_1 \, {\rm arctanh} \, \frac{2\sqrt{b^2- \omega_1^2}}{a^2}  \hspace{0.5cm},\hspace{0.5cm} Q_2[ \overline{{\cal B}}_1(x) ] = 0 \label{Q01}
\end{equation}
whereas the total energy is
\begin{equation}
E[\overline{{\cal B}}_1(x )] = \frac{1}{4} a^2 \sqrt{b^2-\omega_1^2} - \frac{Q_1[\overline{{\cal B}}_1(x)]}{8 \omega_1} \Big( a^4-4 b^2-4 \, \omega_1^2  \Big) \hspace{0.4cm} . \label{enerQ1}
\end{equation}

\item \textit{$\widehat{{\cal B}}_1(x)$-type $Q$-balls:} On the other hand, a $|\psi|^6$-model is found when $\phi=0$. The effective potential (\ref{potential06}) in this case is given by
\begin{equation}
\overline{U}(0,f_2; \sigma, a ,b,\omega_1,\omega_2) =\frac{1}{2} \Big( f_2^6 - (a^2 - 2 \,\sigma^2) f_2^4 + [b^2  +  \sigma^2 (\sigma^2-a^2)] f_2^2 \Big) - \frac{1}{2} \, \omega_2^2 \, f_2^2 \hspace{0.4cm}. \label{pot02}
\end{equation}
The expression (\ref{pot02}) follows the same functional form that (\ref{pot01}). This can be checked by redefining the parameters
\begin{equation}
a_1^2 = a^2 - 2\sigma^2 \hspace{0.5cm} , \hspace{0.5cm} b_1^2 = b^2 - \sigma^2(a^2- \sigma^2) \label{newpara}
\end{equation}
and rewriting the expression of the restricted effective potential (\ref{pot02}) in terms of these new parameters,
\[
\overline{U}(0,f_2) =\frac{1}{2} \Big( f_2^6 - a_1^2 f_2^4 + b_1^2  f_2^2 \Big) - \frac{1}{2} \, \omega_2^2 \, f_2^2 \hspace{0.4cm}.
\]
We can take advantage of this fact to directly obtain the expressions that characterize the $\widehat{{\cal B}}_1(x)$ solutions. Now, for example, the Hessian operator along the $\psi$-direction evaluated at the origin of the internal plane is
\[
\frac{\partial^2 \overline{U}(0,f_2)}{\partial f_2^2} = b_1^2  -\omega_2^2 = b^2- \sigma^2(a^2-\sigma^2) -\omega_2^2
\]
whereas the zeroes of the potential (\ref{pot02}) along the $|\psi|$-axis are
\begin{equation}
\widetilde{f}_2 =  \pm \frac{1}{\sqrt{2}} \Big( a_1^2 \pm \sqrt{a_1^4 - 4 (b_1^2 - \omega_2^2)} \,\, \Big)^\frac{1}{2}  =\pm \frac{1}{\sqrt{2}} \Big( a^2 - 2\sigma^2 \pm \sqrt{a^4 - 4 (b^2 -  \omega_2^2)} \,\, \Big)^\frac{1}{2} \hspace{0.4cm}. \label{roots02}
\end{equation}
To obtain real roots from (\ref{roots02}) we impose the additional condition
\begin{equation}
a^2> 2 \, \sigma^2 \hspace{0.4cm}. \label{condition03}
\end{equation}
The constraints on the internal rotation frequencies of these $Q$ balls now read
\begin{equation}
b^2 -\frac{1}{4} a^4 \leq \, \omega_2^2  \, \leq b^2 - \sigma^2(a^2-\sigma^2)\hspace{0.4cm},\label{rango02}
\end{equation}
and for these frequencies the solutions can be written as
\begin{equation}
\widehat{{\cal B}}_1(x;\omega_2) =(f_1(x),f_2(x)) = \left(0, \frac{\sqrt{2(b_1^2-\omega_2^2)}}{\sqrt{a_1^2 + \sqrt{a_1^4- 4(b_1^2-\omega_2^2)} \cosh (2\sqrt{b_1^2-\omega_2^2} \, \overline{x}) }} \right)  \hspace{0.4cm}. \label{qball02}
\end{equation}
In this case, the Noether charges are
\begin{equation}
Q_1[\widehat{{\cal B}}_1(x)]= 0 \hspace{0.5cm},\hspace{0.5cm}  Q_2[\widehat{{\cal B}}_1(x)]= \omega_2 \, {\rm arctanh} \frac{2\sqrt{b^2-\sigma^2(a^2-\sigma^2)-\omega_2^2}}{a^2-2\sigma^2}  \label{Q02}
\end{equation}
with a total energy
\begin{equation}
E [\widehat{{\cal B}}_1(x)] = \frac{1}{4} (a^2-2\sigma^2) \sqrt{b^2-\sigma^2(a^2-\sigma^2)-\omega_2^2} - \frac{Q_2[\widehat{{\cal B}}_1(x)]}{8 \omega_2} \Big(a^4- 4 b^2 -4 \omega_2^2) \Big) \hspace{0.4cm}. \label{enerQ2}
\end{equation}

\end{itemize}

\begin{figure}[h]
	\centerline{\includegraphics[height=3.cm]{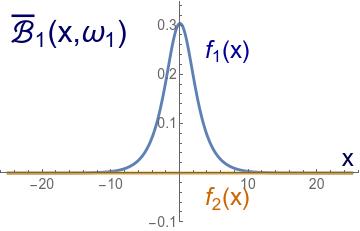} \hspace{1cm} \includegraphics[height=3.cm]{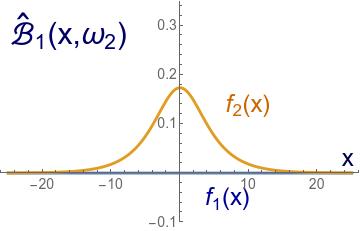} \hspace{1cm} \includegraphics[height=3.cm]{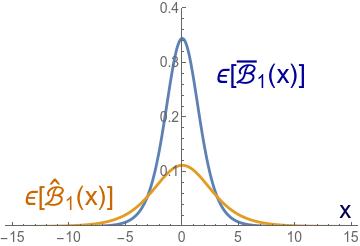}}
	\caption{\small Profiles of the components $f_i(x)$ for the $\overline{{\cal B}}_1(x)$ (left) and $\widehat{{\cal B}}_1(x)$ (middle) solutions for the parameter values $\sigma=0.25$, $a=1.75$, $b=2.0$ and $\omega_1=\omega_2=1.93$. Energy densities (right) for the previous solutions.} \label{fig:profilef}
\end{figure}

\begin{figure}[h]
	\centerline{\includegraphics[height=3.cm]{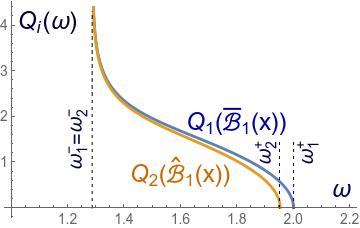} \hspace{1cm} \includegraphics[height=3.cm]{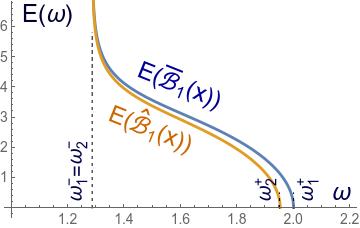}}
	\caption{\small Graphics of the conserved Noether charges $Q_i$ (left) and the total energies $E$ (right) of the two types of single $Q$-balls as a function of $\omega$ for the parameter values $\sigma=0.25$, $a=1.75$ and $b=2.0$.} \label{fig:profile}
\end{figure}

Curiously, the lowest value of the frequency range defining the two types of $Q$ balls coincides, see (\ref{rango01}) and (\ref{rango02}), that is, $\omega_2^{-}=\omega_1^{-}$. On the other hand, the highest internal rotation frequency for the $\widehat{{\cal B}}_1(x)$-type $Q$ balls is smaller than the value found for the $\overline{{\cal B}}_1(x)$ solutions, that is, $\omega_2^{+} \leq \omega_1^{+}$. In Figure \ref{fig:profilef}(a) and (b) the profiles of the functions $f_i(x)$ are respectively depicted for the two types of $Q$-balls for the specific choice of the model parameters $\sigma=0.25$, $a=1.75$ and $b=2.0$. We consider these values as representative of the model. The behavior of the solutions is completely similar for other values of the model parameters. For our choice the internal rotational frequencies are approximately restricted to the values $1.28657\leq \omega_1 \leq 2.0$ and $1.28657\leq \omega_2 \leq 1.95256$. In Figure \ref{fig:profilef}(c) it can be checked that these solutions consist of only one energy lump, that is, they are single $Q$-balls. In particular, the energy density of the first type of these non-topological solitons is more concentrated and compact than the energy density of the second type when the same internal rotational frequency $\omega$ is considered. Note that, in general, ${\cal E}[\overline{{\cal B}}_1(0;\omega)] < {\cal E}[\widehat{{\cal B}}_1(0;\omega)]$ and the size of the $Q$-balls can be estimated respectively as
\[
\Delta \overline{{\cal B}}_1(x;\omega)= \frac{1}{\sqrt{b^2-\omega^2} } \hspace{0.4cm},\hspace{0.4cm} \Delta \widehat{{\cal B}}_1(x;\omega)= \frac{1}{\sqrt{b^2-\sigma^2(a^2-\sigma^2)-\omega^2}}
\]
such that $\Delta \overline{{\cal B}}_1(x;\omega)\leq \Delta \widehat{{\cal B}}_1(x;\omega)$. In Figure \ref{fig:profile} the Noether charges and the total energies for the two types of $Q$-balls are plotted as a function of the internal rotational frequency $\omega$. It can be checked that $Q_2 [\widehat{{\cal B}}_1(x;\omega)] \leq Q_1 [\overline{{\cal B}}_1(x;\omega)]$ and $E [\widehat{{\cal B}}_1(x;\omega)] \leq E [\overline{{\cal B}}_1(x;\omega)]$. If the Noether charges of these two solutions are compared when the asymmetry parameter $\sigma$ is small while rotating with the same frequency $\omega$, the relation
\begin{equation}
\Delta Q =  Q_1 [\overline{{\cal B}}_1(x;\omega)] - Q_2 [\widehat{{\cal B}}_1(x;\omega)]= \sigma^2 \frac{\omega}{\sqrt{b^2-\omega^2}} + O(\sigma^4) \label{Qdiff}
\end{equation}
is found. From (\ref{Qdiff}) it can be noted that the difference $\Delta Q$ between the Noether charges for the two types of $Q$-balls depends on $\sigma^2$, which means that the two charges $Q_1 [\overline{{\cal B}}_1(x;\omega)]$ and $Q_2 [\widehat{{\cal B}}_1(x;\omega)]$ are approximately equal for a large range of small values of $\sigma$.

\subsection{Composite $Q$-Balls }

In this section we shall investigate the existence of $Q$-balls with two non-null complex scalar fields. Using the mechanical analogy these solutions could be interpreted as a particle which asymptotically leaves the origin of the internal $(f_1,f_2)$-plane, travels in a bounded region of this plane and asymptotically returns to the origin. A profitable method employed in the identification of solutions in some types of deformations of $O(N)$ invariant models consists of exploring the separability of the model when elliptic variables are used in the internal space. Bearing this in mind, we introduce the elliptic coordinates in the form
\begin{equation}
\xi_\pm^* f_1= \frac{1}{\Omega} \, u \, v \hspace{1cm}, \hspace{1cm} \xi_\pm^*  f_2 = \pm \frac{1}{\Omega} \sqrt{(u^2-\Omega^2)(\Omega^2-v^2)} \label{elipticas}
\end{equation}
where $u\in [\Omega,+\infty)$ and $v\in [-\Omega,\Omega]$. The first condition for our model to be separable in elliptic variables is to set the eccentricity parameter $\Omega$ in (\ref{elipticas}) to the deformation parameter $\sigma$ arising in the potential (\ref{potential03}), i.e.,
\[
\Omega= \sigma  \hspace{0.4cm}.
\]
This implies that the effective potential (\ref{potential06}) can be expressed in the new coordinates as
\begin{equation}
\xi_\sigma^* \overline{U}(f_1,f_2) = \frac{1}{u^2-v^2} \Big[ f(u) + g(v) + \frac{\omega_2^2 - \omega_1^2}{2\sigma^2} \, u^2v^2 (u^2 -v^2) \Big] \label{potelip}
\end{equation}
where
\begin{eqnarray}
f(u) &=& \frac{1}{2} \Big[ u^2 (u^4-a^2 u^2+ b^2 - \omega_2^2) (u^2- \sigma^2) \Big]  \hspace{0.4cm},\label{fu} \\
g(v) &=& \frac{1}{2} \Big[ v^2 (v^4-a^2 v^2+ b^2 - \omega_2^2) (\sigma^2 - v^2) \Big] \hspace{0.4cm}. \label{gv}
\end{eqnarray}
From (\ref{potelip}) it can be checked that separability is attained if and only if the internal rotation frequencies around each of the complex fields are equal, that is, $\omega_1=\omega_2=\omega$. In this Section, we shall investigate this class of solutions and we shall show that they involve a very rich structure. Obviously, the possible values of $\omega$ are restricted to the intersection of the allowed values for $\omega_1$ and $\omega_2$. The effective Lagrangian (\ref{reducedLagran}) can be written in elliptic variables as
\[
\overline{E}[\xi^* f_1,\xi^* f_2] = \frac{1}{2} \frac{u^2-v^2}{u^2 -\sigma^2} \Big( \frac{du}{dx}\Big)^2 +  \frac{1}{2} \frac{u^2-v^2}{\sigma^2 - v^2} \Big( \frac{dv}{dx}\Big)^2  + \frac{f(u)+g(v)}{u^2-v^2}\hspace{0.4cm},
\]
which allows us to define the generalized momenta
\[
p_u = \frac{\partial {\cal L}}{\partial \left( \frac{du}{dx} \right)} = \frac{u^2 - v^2}{u^2- \sigma^2} \, \frac{du}{dx} \hspace{1cm},\hspace{1cm} p_v = \frac{\partial {\cal L}}{\partial \left( \frac{dv}{dx} \right)} = \frac{u^2 - v^2}{\sigma^2- v^2} \, \frac{dv}{dx} \hspace{0.4cm}.
\]
Now, the Hamiltonian associated to our problem reads
\[
{\cal H} = \frac{1}{u^2-v^2} (h_u+h_v)
\]
where
\[
h_u=\frac{1}{2} (u^2-\sigma^2)p_u^2 -f(u) \hspace{1cm} \mbox{and} \hspace{1cm} h_v=\frac{1}{2} (\sigma^2-v^2)p_v^2 -g(v) \hspace{0.4cm}.
\]
The Hamilton-Jacobi equation
\begin{equation}
\frac{\partial {\cal J}}{\partial x} + {\cal H} \Big( \frac{\partial {\cal J}}{\partial u}, \frac{\partial {\cal J}}{\partial v},u,v \Big) = 0 \label{HJequation}
\end{equation}
is now completely separable. In order to abbreviate the formulas we shall use \lq prime\rq-notation to stand for derivatives with respect to the space coordinate $x$. For example, $u'= \frac{du}{dx}$ and $v'=\frac{dv}{dx}$. If the separation ansatz for Hamilton's principle function ${\cal J} = {\cal J}_x(x) + {\cal J}_u(u) + {\cal J}_v(v)$ is substituted into (\ref{HJequation}) the relation
\[
\frac{1}{2} (u^2-\sigma^2) ({\cal J}_u ')^2 - E u^2-f(u)=F = -\frac{1}{2} (\sigma^2-v^2) ({\cal J}_v ')^2 - E v^2 + g(v)
\]
holds. The solutions of the PDE (\ref{HJequation}) can be expressed in terms of the solutions of the ODEs
\[
{\cal J}_x ' = -E \hspace{0.4cm},\hspace{0.4cm} {\cal J}_u' = {\rm sign}(u') \sqrt{\frac{2(F+Eu^2+f(u))}{u^2-\sigma^2}} \hspace{0.4cm},\hspace{0.4cm} {\cal J}_v'= {\rm sign}(v') \sqrt{\frac{2(-F-E v^2+g(v))}{\sigma^2- v^2}}
\]
which leads to
\begin{eqnarray*}
{\cal J}_x &=& -E x  \hspace{0.4cm},\\
{\cal J}_u &=&  {\rm sign}(u') \int dx \sqrt{\frac{2(F+Eu^2+f(u))}{u^2-\sigma^2}}  \hspace{0.4cm},\\
{\cal J}_v &=& {\rm sign}(v') \int dx \sqrt{\frac{2(-F-E v^2+g(v))}{\sigma^2- v^2}}  \hspace{0.4cm}.
\end{eqnarray*}
Once Hamilton's principle function ${\cal J}$ has been obtained the solutions can be determined from the relations
\begin{equation}
\frac{\partial {\cal J}}{\partial F} =\gamma_1 \hspace{0.5cm},\hspace{0.5cm} \frac{\partial {\cal J}}{\partial E} =\gamma_2 \label{HJsolutions}
\end{equation}
where $\gamma_1,\gamma_2\in \mathbb{R}$ identify every solution of the problem. The first equation in (\ref{HJsolutions}) provides the trajectories or orbits of the solutions in the $(f_1,f_2)$-plane parametrized by the value of $\gamma_1$ whereas the second equation in (\ref{HJsolutions}) specifies the dependence of the solutions with respect to the space $x$. Taking into account that the asymptotic conditions (\ref{asymptotics}) implies that $E=F=0$ the relations (\ref{HJsolutions}) lead to the quadratures
\begin{eqnarray}
	&& {\rm sign}(u') \int \frac{du}{\sqrt{(u^2-\sigma^2) f(u)}} - {\rm sign}(v') \int \frac{dv}{\sqrt{(\sigma^2-v^2) g(v)}} = \sqrt{2} \, \gamma_1  \hspace{0.4cm} , \label{HJsol01} \\
	&& {\rm sign}(u') \int \frac{u^2 \, du}{\sqrt{(u^2-\sigma^2) f(u)}} - {\rm sign}(v') \int \frac{v^2 \,dv}{\sqrt{(\sigma^2-v^2) g(v)}} = \sqrt{2} \, (x+\gamma_2) \hspace{0.4cm} . \label{HJsol02}
\end{eqnarray}
For our model the expressions (\ref{HJsol01}) and (\ref{HJsol02}) become
\begin{eqnarray*}
	&& {\rm sign}(u') \int \frac{du}{u (u^2-\sigma^2) \sqrt{u^4-a^2u^2+b^2-\omega^2}} - {\rm sign}(v') \int \frac{dv}{ v (\sigma^2-v^2) \sqrt{v^4-a^2v^2 + b^2-\omega^2}} =  \, \gamma_1  \hspace{0.2cm} , \\
	&& {\rm sign}(u') \int \frac{u \, du}{(u^2-\sigma^2) \sqrt{u^4-a^2u^2+b^2-\omega^2}} - {\rm sign}(v') \int \frac{v \,dv}{(\sigma^2-v^2) \sqrt{v^4-a^2v^2 + b^2-\omega^2}} =  \, x+\gamma_2 \hspace{0.2cm} ,
\end{eqnarray*}
where we recall that $\omega=\omega_1=\omega_2$ is the synchronized frequency around the complex coordinate axes at which the solutions we are looking for are spinning, that is, $\phi(x,t)=f_1(x) e^{i\omega t}$ and $\psi(x,t)=f_2(x) e^{i\omega t}$. Note that the parameter $\gamma_2$ only represents a translation of the solution along the $x$-axis. In order to alleviate the notation in the calculations of the previous quadratures we shall denote the roots of the polynomial
\begin{equation}
p(z)=z^2-a^2 z+b^2-\omega^2 \label{polinop}
\end{equation}
as
\begin{equation}
r_{\pm} = \frac{1}{2} \Big( a^2 \pm \sqrt{a^4 - 4 (b^2 - \omega^2)} \Big) \hspace{0.4cm} .  \label{roots}
\end{equation}
Clearly, the inequation $0 \leq v^2 \leq \sigma^2 \leq u^2\leq r_- \leq r_+$ holds for the solutions which we are looking for. If we define
\begin{equation}
P(z)= \sqrt{(z-r_-)(z-r_+)} \hspace{0.5cm} , \hspace{0.5cm} R(z) = \sqrt{\frac{r_+-z}{r_--z}} \label{PyQ}
\end{equation}
the expression of the $Q$-ball orbits in the $(f_1,f_2)$-space is given by
\begin{equation}
\left[ \left| \frac{R(u^2)-R(0)}{R(u^2)+R(0)} \right|^{\frac{P(\sigma^2)}{P(0)} }  \left| \frac{R(u^2)+R(\sigma^2) }{R(\sigma^2) - R(u^2) } \right| \right]^{{\rm sign}(u')}  \left[ \left| \frac{R(v^2)-R(0)}{R(v^2)+R(0)} \right|^{\frac{P(\sigma^2)}{P(0)} }  \left| \frac{R(v^2)+R(\sigma^2) }{R(v^2)-R(\sigma^2) } \right| \right]^{{\rm sign}(v')} = e^{2\sigma^2 P(\sigma^2) \gamma_1} \label{orb1}
\end{equation}
while the spatial dependence of these solutions can be obtained from the relation
\begin{equation}
\left| \frac{R(u^2)+R(\sigma^2) }{R(\sigma^2) - R(u^2)} \right|^{{\rm sign}(u')}  \left| \frac{R(v^2)+R(\sigma^2) }{R(v^2)-R(\sigma^2) } \right|^{{\rm sign}(v')}   = e^{2P(\sigma^2) \, \overline{x}} \label{orb2}
\end{equation}
where $\overline{x}=x+\gamma_2$. It can be checked that relations (\ref{orb1}) and (\ref{orb2}) are invariant under the transformations $\gamma_1 \rightarrow -\gamma_1$ and $\overline{x} \rightarrow -\overline{x}$. This means that solutions with $\gamma_1<0$ can be constructed from solutions with $\gamma_1>0$ by simply flipping the space axis $x$. Plugging (\ref{orb2}) into (\ref{orb1}) the following more simple relation
\begin{equation}
\left| \frac{R(u^2)-R(0)}{R(u^2)+R(0)} \right|^{{\rm sign}(u')}  \left| \frac{R(v^2)-R(0)}{R(v^2)+R(0)} \right|^{{\rm sign}(v')} = e^{2 P(0) (\sigma^2 \gamma_1 -\overline{x})} \label{orb3}
\end{equation}
is deduced.

\begin{figure}[h]
	\centerline{\begin{tabular}{|c||ccc|} \hline
			& {\Large$\gamma_1=0$} & {\Large$\gamma_1$ intermediate} & {\Large$\gamma_1$ large} \\ \hline\hline
			\rotatebox{90}{\Large\hspace{0.6cm}$\omega=1.3$} &	\includegraphics[height=3cm]{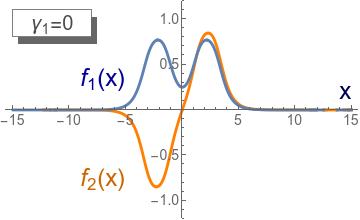} & \includegraphics[height=3cm]{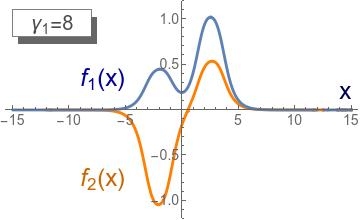}  & \includegraphics[height=3cm]{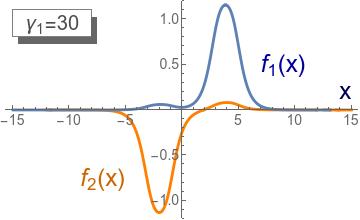} \\
			\rotatebox{90}{\Large\hspace{0.6cm}$\omega=1.7$} &	\includegraphics[height=3cm]{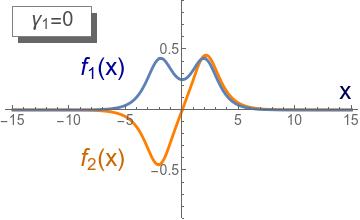} & \includegraphics[height=3cm]{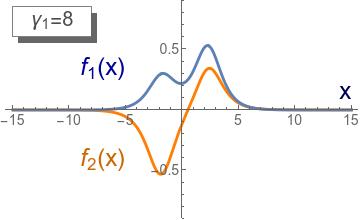}  & \includegraphics[height=3cm]{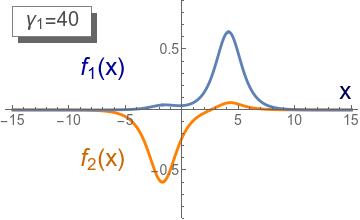} \\
			\rotatebox{90}{\Large\hspace{0.6cm}$\omega=1.9$} &	\includegraphics[height=3cm]{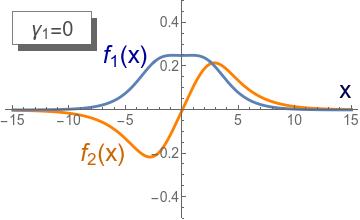} & \includegraphics[height=3cm]{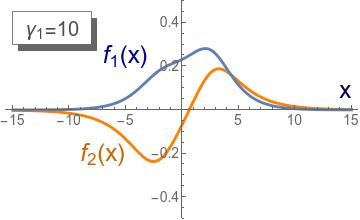}  & \includegraphics[height=3cm]{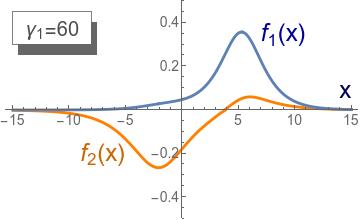} \\
			\rotatebox{90}{\Large\hspace{0.6cm}$\omega=1.93$} &	\includegraphics[height=3cm]{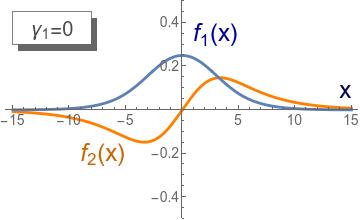} & \includegraphics[height=3cm]{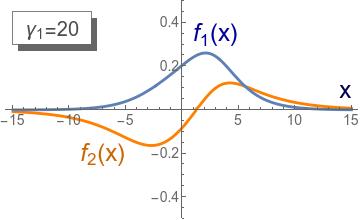}  & \includegraphics[height=3cm]{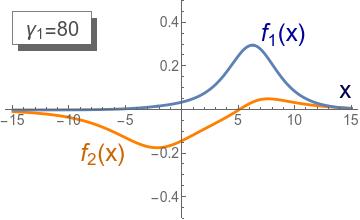} \\ \hline
	\end{tabular}}
	\caption{Graphics of the profiles $f_1$ (blue curve) and $f_2$ (orange curve) of the composite ${\cal B}_{2}(x;\omega,\gamma_1)$-balls for several values of the internal rotational frequency $\omega$ and the family parameter $\gamma_1$. The model parameters have been set as $\sigma=0.25$, $a=1.75$ and $b=2.0$.} \label{fig:profileComp}
\end{figure}

The set of equations (\ref{orb2}) and (\ref{orb3}) can be used to obtain an analytical expression of the families of $Q$-balls. Although it is a lengthy expression, it is worthwhile (from our point of view) to write it down in a sequential way because it can be used to plot the profiles of $Q$-balls as well as its energy densities, conserved Noether charges, etc. The equations (\ref{orb2}) and (\ref{orb3}) can be solved for the functions $R(u^2)$ and $R(v^2)$, which in turn provide us with the profile of the elliptic variables for the $Q$-balls in the form
\begin{equation}
u (x; \sigma,a,b,\omega ) = \sqrt{ \frac{r_+ -  r_-(S_1-S_2)^2}{1-(S_1-S_2)^2}} \hspace{0.5cm},\hspace{0.5cm} v(x; \sigma,a,b,\omega ) = {\rm sign}(v) \sqrt{ \frac{r_+ -  r_-(S_1+S_2)^2}{1-(S_1+S_2)^2}} \label{profileuv}
\end{equation}
where
\begin{equation}
S_1= \frac{E_2 [R^2(\sigma^2)-R^2(0)]}{2(R(0) + E_1 E_2 R(\sigma^2))}  \hspace{0.5cm}, \hspace{8.5cm} \label{profileuv1}
\end{equation}
\begin{equation}
S_2=  \frac{\sqrt{4R(0) R(\sigma^2) [R(\sigma^2)+E_1 E_2 R(0)][R(0)+E_1 E_2 R(\sigma^2)]+ E_2^2 [R^2(\sigma^2)-R^2(0)]^2}}{2(R(0) + E_1 E_2 R(\sigma^2))} \label{profileuv2}
\end{equation}
and
\[
E_1  =  \tanh [P(\sigma^2) \overline{x}] \hspace{1cm},\hspace{1cm} E_2 =  \tanh [P(0) (\sigma^2 \gamma_1 - \overline{x})] \hspace{0.4cm} .
\]

\begin{figure}[H]
	\centerline{\begin{tabular}{|c||ccc|} \hline
			& {\Large$\gamma_1=0$} & {\Large$\gamma_1$ intermediate} & {\Large$\gamma_1$ large} \\ \hline\hline
			\rotatebox{90}{\Large\hspace{1.2cm}$\omega=1.3$} &	\includegraphics[width=3.cm]{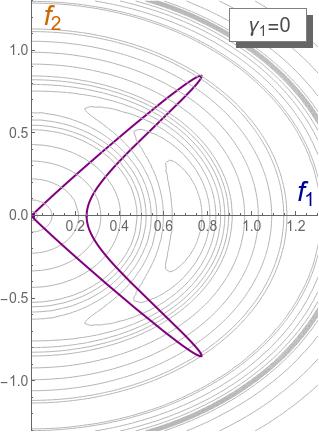} & \hspace{0.5cm} \includegraphics[width=3.cm]{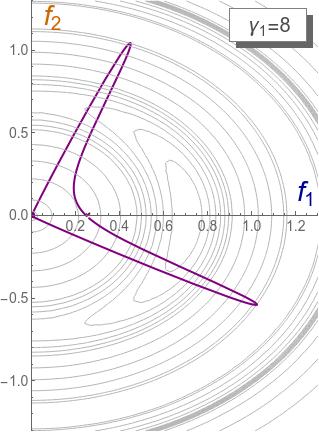} \hspace{0.5cm} & \includegraphics[width=3.cm]{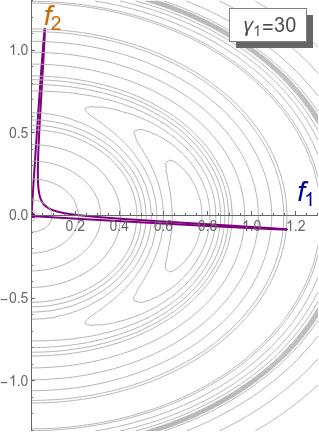} \\ [0.3cm]
			\rotatebox{90}{\Large\hspace{1.2cm}$\omega=1.7$} &	\includegraphics[width=3.cm]{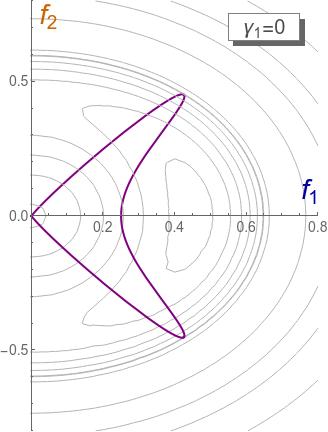} & \hspace{0.5cm} \includegraphics[width=3.cm]{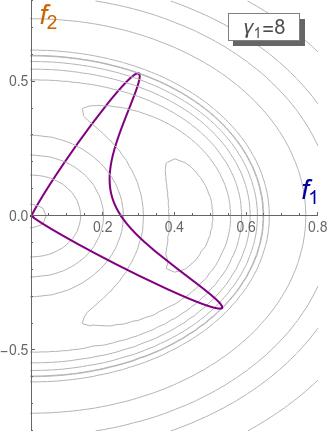} \hspace{0.5cm} & \includegraphics[width=3.cm]{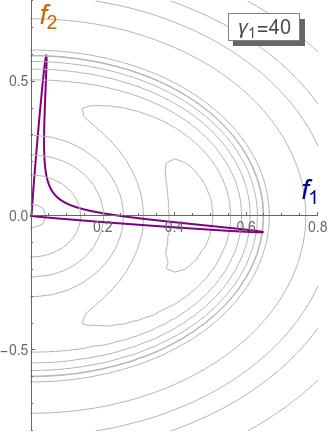} \\  [0.3cm]
			\rotatebox{90}{\Large\hspace{1.2cm}$\omega=1.9$} &	\includegraphics[width=3.cm]{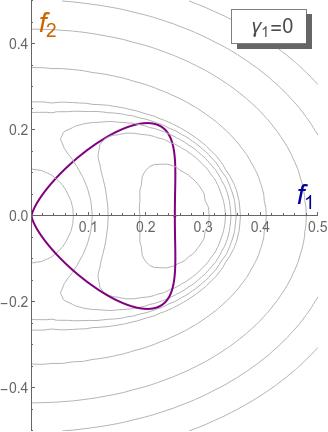} & \hspace{0.5cm} \includegraphics[width=3.cm]{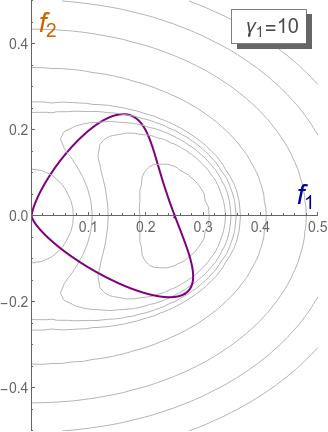} \hspace{0.5cm} & \includegraphics[width=3.cm]{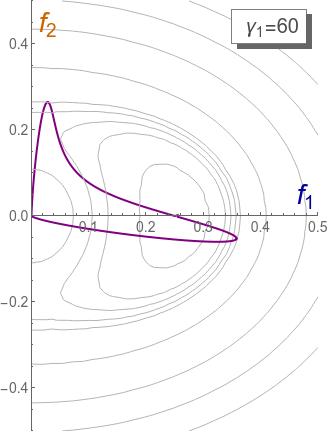} \\  [0.3cm]
			\rotatebox{90}{\Large\hspace{1.2cm}$\omega=1.93$} &	\includegraphics[width=3.cm]{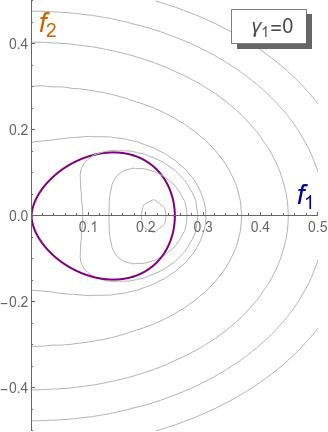} & \hspace{0.5cm} \includegraphics[width=3.cm]{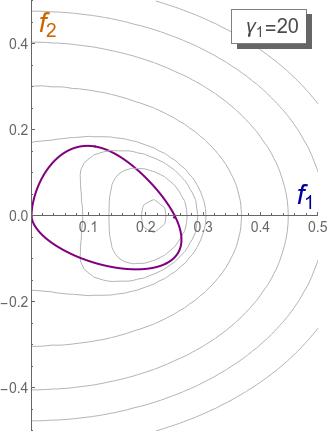} \hspace{0.5cm} & \includegraphics[width=3.cm]{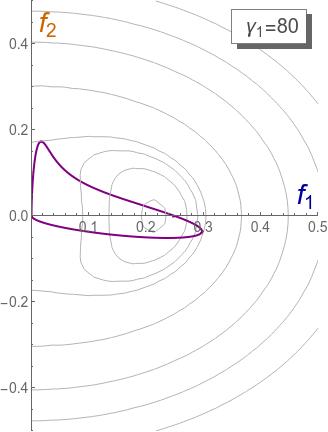} \\ \hline
	\end{tabular}}
	\caption{Graphics of the ${\cal B}_2(x;\omega,\gamma_1)$-orbits (purple curve) for several values of the internal rotational frequency $\omega$ and the family parameter $\gamma_1$. The model parameters have been set as $\sigma=0.25$, $a=1.75$ and $b=2.0$. A contour plot for the effective potential density $\overline{U}(f_1,f_2)$ has been used in the previous graphics.} \label{fig:OrbitsComp}
\end{figure}

The profile of the original variables $f_1$ and $f_2$ can be obtained from (\ref{profileuv}) by using the change of variables (\ref{elipticas}). We shall refer to these solutions as ${\cal B}_2(x;\omega,\gamma_1)$. In Figure \ref{fig:profileComp} the profiles of the functions $f_1$ and $f_2$ associated with these $Q$-balls (determined by the expressions (\ref{profileuv}), (\ref{profileuv1}) and (\ref{profileuv2})) have been plotted for several values of the orbit parameter $\gamma_1$ assuming the representative model parameters $\sigma=0.25$, $a=1.75$ and $b=2.0$ chosen in this paper. In this case the internal rotational frequencies $\omega$ take values on the interval $\omega\in [1.28657,1.95256]$. The profiles of the ${\cal B}_2(x;\omega,\gamma_1)$-solutions have also been depicted for several values of these frequencies $\omega$. Indeed, Figure \ref{fig:profileComp} has been arranged in tabular form where rows are associated with a particular value of the frequency $\omega$ and columns correspond to different values of the orbit parameter $\gamma_1$. Three different types of values of this parameter are considered in order to show the distinct behaviors of this family of $Q$-balls: in the first column $\gamma_1$ is set to zero while in the third one a large value of $\gamma_1$ is fixed (whose particular value depends on the rotational frequency $\omega$). In the second column an intermediate value between the previous ones is considered. This categorization of the values of $\gamma_1$ is related to the gradual loss of the symmetry/antisymmetry in the components $f_1$ and $f_2$. For $\gamma_1=0$ the function $f_1$ is an even configuration while $f_2$ is odd. As the orbit parameter $\gamma_1$ increases this symmetry breaks down, see the middle column in Figure \ref{fig:profileComp}. However, the most remarkable feature of the ${\cal B}_2(x;\omega,\gamma_1)$-family can be observed when the orbit parameter $\gamma_1$ is large enough, see the third column in Figure \ref{fig:profileComp}. In this case the profiles of the components $f_1$ and $f_2$ decouple each other. They are significantly different from zero for regions where the other component almost vanishes. The resulting configurations for each field correspond to the profiles of a single $\overline{\cal B}_1(x;\omega)$-type $Q$-ball and a single $\widehat{\cal B}_1(x;\omega)$-type $Q$-ball. This suggests that the ${\cal B}_2(x;\omega,\gamma_1)$ solutions are non-linear combinations of the two types of single $Q$-balls where the orbit parameter $\gamma_1$ measures the distance between them. This justifies the subscript 2 included in the notation for these solutions. In particular, when $|\gamma_1| \rightarrow \infty$ the two single $Q$-balls follow the expressions (\ref{qball01}) and (\ref{qball02}) and they are infinitely separated. For $\gamma_1=0$ they are completely overlapped.

The previous remark is also clearly observed in Figure \ref{fig:OrbitsComp}, where the orbits associated to the composite ${\cal B}_2(x;\omega,\gamma_1)$-solutions are displayed in the internal space $(f_1,f_2)$. For the sake of comparison we have employed in these graphics the same values of $\gamma_1$ and $\omega$ introduced in Figure \ref{fig:profileComp}. Using the previously mentioned mechanical analogy, the ${\cal B}_2(x;\omega,\gamma_1)$-orbit shown in the first plot in Figure \ref{fig:OrbitsComp} (for $\omega=1.3$ and $\gamma_1=0$) can be interpreted as the trajectory of a particle asymptotically leaving the maximum located at the origin of the plane $(f_1,f_2)$ in a certain direction. The particle travels up to a point where the potential barrier forces the particle to move backwards approaching to the $f_1$-axis and crossing the focus point $F=(\sigma,0)$. At this point the particle is again expelled far away from the origin. A second bounce takes place which makes the particle approach again to the origin, which now is asymptotically reached. This behavior is replicated in the rest of the cases. As previously mentioned the orbits with $\gamma_1=0$ are symmetric with respect to reflections $f_2\rightarrow -f_2$. This symmetry is lost for non-null values of $\gamma_1$. Indeed, for $|\gamma_1|\rightarrow \infty$ the orbit becomes the union of two segments, one along the axis $f_1$ and the other along the axis $f_2$, which correspond to the single $Q$-ball orbits. This behavior corroborates the previous claim that the composite ${\cal B}_2(x;\omega,\gamma_1)$ solutions consist of two single $\overline{\cal B}_1(x;\omega)$ and $\widehat{\cal B}_1(x;\omega)$ lumps.

\begin{figure}[h]
	\centerline{\begin{tabular}{|c||ccc|} \hline
			& {\Large$\gamma_1=0$} & {\Large$\gamma_1$ intermediate} & {\Large$\gamma_1$ large} \\ \hline\hline
			\rotatebox{90}{\Large\hspace{0.6cm}$\omega=1.3$} &	\includegraphics[height=3cm]{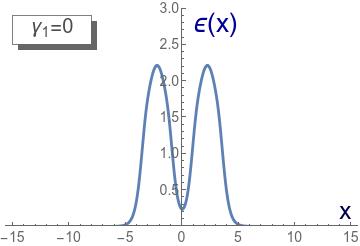} & \includegraphics[height=3cm]{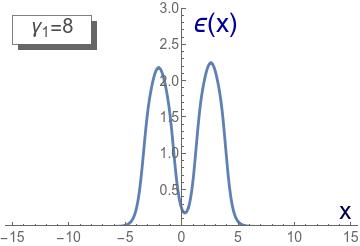}  & \includegraphics[height=3cm]{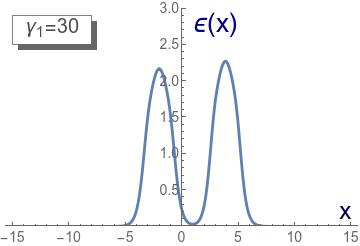} \\
			\rotatebox{90}{\Large\hspace{0.6cm}$\omega=1.7$} &	\includegraphics[height=3cm]{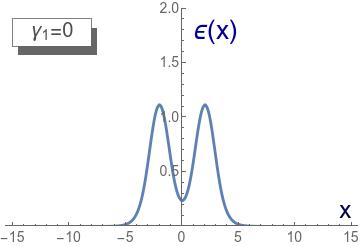} & \includegraphics[height=3cm]{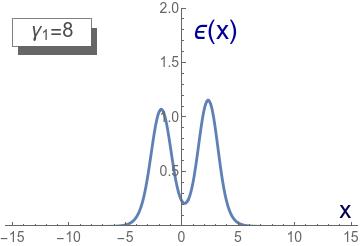}  & \includegraphics[height=3cm]{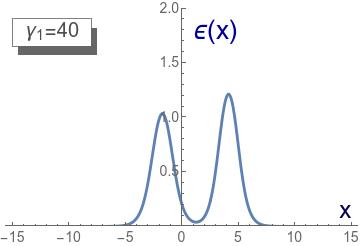} \\
			\rotatebox{90}{\Large\hspace{0.6cm}$\omega=1.9$} &	\includegraphics[height=3cm]{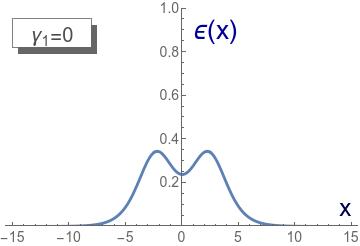} & \includegraphics[height=3cm]{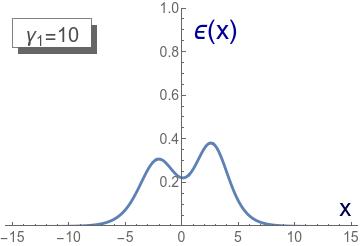}  & \includegraphics[height=3cm]{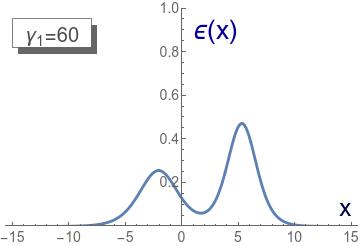} \\
			\rotatebox{90}{\Large\hspace{0.6cm}$\omega=1.93$} &	\includegraphics[height=3cm]{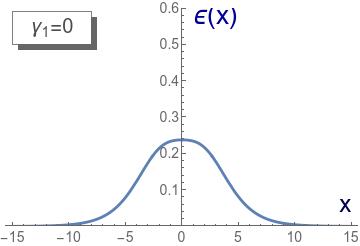} & \includegraphics[height=3cm]{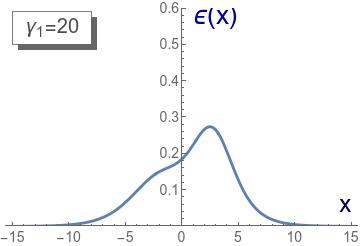}  & \includegraphics[height=3cm]{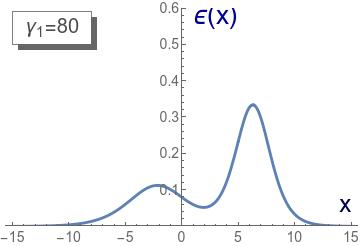} \\ \hline
	\end{tabular}}
	\caption{Graphics of the energy density of the ${\cal B}_2(x;\omega,\gamma_1)$ solutions for several values of the internal rotational frequency $\omega$ and the family parameter $\gamma_1$. The model parameters have been set as $\sigma=0.25$, $a=1.75$ and $b=2.0$.} \label{fig:EnerComp}
\end{figure}

Finally, Figure \ref{fig:EnerComp} illustrates the energy density of the solutions plotted in Figure \ref{fig:profileComp}. As expected, when the orbit parameter $\gamma_1$ is large enough two different energy lumps are discerned for any value of the rotational frequency $\omega$, each one associated to the previously mentioned single $Q$-balls. Note that almost for all the values of the internal rotational frequencies $\omega$ these two energy lumps arise even for $\gamma_1=0$ where the symmetric configuration takes place. Only for very high values of $\omega$ the two lumps merge for small values of the orbit parameter $\gamma_1$. If $\gamma_1$ is large and negative the solution (\ref{profileuv}) describes a single $\overline{\cal B}_1(x;\omega)$ lump placed to the left of a single $\widehat{\cal B}_1(x;\omega)$ solution, which are well separated. As $\gamma_1$ increases the two single $Q$-balls approach each other while losing its identity and eventually giving place to two identical lumps when $\gamma=0$. In this situation the two lumps remains separated by a certain distance. However, as the value of $\gamma_1$ asymptotically increases the lump on the left gradually becomes a $\widehat{\cal B}_1(x;\omega)$ lump while the one on the right becomes a $\overline{\cal B}_1(x;\omega)$-type $Q$-ball.

There are another two interesting properties of the family of composite $Q$-balls introduced in this section. It can be proved that:
\begin{enumerate}
	\item The ${\cal B}_2(x;\omega,\gamma_1)$-family is energy degenerate and complies with the energy sum rule
	\begin{equation}
	E[{\cal B}_2(x;\omega,\gamma_1)] = E[\overline{\cal B}_1(x;\omega)]+E[\widehat{\cal B}_1(x;\omega)] \label{SumRuleE}
	\end{equation}
	that is, the energy of the composite $Q$-balls is equal to the sum of the energies of the two types of single $Q$-balls.
	
	\item The sum of the two $Q_i$-charges is equal for every member of the family of composite $Q$-balls, which means that $Q_1[{\cal B}_2(x;\omega,\gamma_1)] + Q_2[{\cal B}_2(x;\omega,\gamma_1)]$ is independent of the family parameter $\gamma_1$. Besides, the relation
	\begin{equation}
	Q_1[{\cal B}_2(x;\omega,\gamma_1)] + Q_2[{\cal B}_2(x;\omega,\gamma_1)] = Q_1[\overline{\cal B}_1(x;\omega)]+Q_2[\widehat{\cal B}_1(x;\omega)] \label{SumRuleQ}
	\end{equation}
	holds.
\end{enumerate}

In order to prove the previous statements, a Bogomolnyi arrangement of the functional (\ref{reducedLagran}) (the analogue mechanical energy functional) written in elliptic variables will be introduced in this framework. By definition, a superpotential expressed in certain generalized coordinates $\{u^i\}$ verifies 
\[
 \frac{1}{2} g^{ij} \frac{\partial W}{\partial u^i} \frac{\partial W}{\partial u^j}   = \overline{U}(u^i)
\]
being $\overline{U}(u^i)$ the effective potential term (\ref{potential06}). In our case, the generalized coordinates are the elliptic coordinates ($u^1=u$, $u^2=v$), the induced metric $g_{ij}$ is given by  $g_{11}=\frac{u^2-v^2}{u^2-\sigma^2}$, $g_{22}=\frac{u^2-v^2}{\sigma^2-v^2}$ and $g_{12}=g_{21}=0$ and the effective potential term $\overline{U}(u,v)$ can be written as $\overline{U}(u,v)=\frac{1}{u^2-v^2}[f(u)+g(v)]$ where $f(u)$ and $g(v)$ were defined respectively in (\ref{fu}) and (\ref{gv}). By taking advantage of the separability of the model the superpotential can be figured out by means of quadratures. It follows the separated form
\begin{equation}
W^{(\alpha,\beta)}(u,v) = (-1)^\alpha W_u (u)+ (-1)^\beta W_v (v)
\label{superpotential01}
\end{equation}
with $\alpha,\beta= 0,1$ and
\begin{equation}
W_z(z)= \frac{2 z^2-a^2}{8} \sqrt{z^4-a^2 z^2 + b^2 - \omega^2} - \frac{a^4-4(b^2-\omega^2)}{16} \log \Big[ a^2 -2  z^2 - 2\sqrt{z^4 - a^2 z^2 + b^2 -\omega^2}  \Big] \label{superpotential02}
\end{equation}
for $z=u,v$. The expression (\ref{superpotential01}) defines four different superpotentials depending on the values of $\alpha$ and $\beta$. Thus, the analogue mechanical energy $\overline{E}$ can be written as
\begin{equation}
\overline{E} [u^i(x)] = \frac{1}{2} \sum_{k=0}^{N-1} \int_{x_k}^{x_{k+1}} \, dx \, g_{ij} \Big[ \frac{du^i}{dx} +  g^{mi} \frac{\partial W^{(\alpha_k,\beta_k)}}{\partial u^m} \Big] \Big[ \frac{du^j}{dx} + g^{nj} \frac{\partial W^{(\alpha_k,\beta_k)}}{\partial u^n} \Big] + T \label{Bogo01}
\end{equation}
where
\begin{equation}
T=\sum_{k=0}^{N-1} \, \Big| W^{(\alpha_k,\beta_k)}[u^i(x_{k+1})] - W^{(\alpha_k,\beta_k)}[u^i(x_k)] \Big| \hspace{0.4cm} . \label{Bogo02}
\end{equation}
The partition $-\infty=x_0 < x_1 < \dots < x_{N-1} < x_N=\infty$ has been introduced in (\ref{Bogo01}) to let us choose different superpotentials for different intervals. $Q$-balls saturate (\ref{Bogo01}) in a piecewise way and verify the first order differential equations
\[
\frac{du^i}{dx} +  g^{mi} \frac{\partial W^{(\alpha_k,\beta_k)}}{\partial u^m} =0 \hspace{0.5cm} \mbox{for} \hspace{0.5cm} x\in [x_k,x_{k+1}]\hspace{0.5cm} \mbox{and} \hspace{0.5cm} k=1,\dots,N \hspace{0.4cm} .
\]
For the $Q$-balls studied in this section the previous equations become
\begin{equation}
\frac{du}{dx} = (-1)^\alpha \frac{\sqrt{u^2-\sigma^2}}{u^2-v^2} \sqrt{2 f(u)} \hspace{0.5cm} , \hspace{0.5cm} \frac{dv}{dx} = (-1)^\beta \frac{\sqrt{\sigma^2-v^2}}{u^2-v^2} \sqrt{2 g(v)} \hspace{0.4cm} , \label{eledo}
\end{equation}
which are completely equivalent to the relations (\ref{HJsol01}) and (\ref{HJsol02}) used previously. The first term in (\ref{Bogo01}) vanishes and the total analogue mechanical energy of these solutions is simply given by $\overline{E}[u^i(x)]= T$. Therefore, $\overline{E}$ is completely determined by the projections of the orbits on each of the elliptic axes. For later use, note that the composite $Q$-balls asymptotically leave the origin $(u,v)=(\sigma,0)$ and travel away up to a turning point $P_1$ where $\frac{du}{dx}=0$, that is, $f(u)=0$. Thus, the maximum value of the $u$-coordinate for these solutions is
\[
u_M= \sqrt{r_-} = \frac{1}{\sqrt{2}} \sqrt{a^2-\sqrt{a^4-4(b^2-\omega^2)}} \hspace{0.4cm}.
\]
This value is independent of the family parameter $\gamma_1$. Subsequently, the orbit crosses through the focus $(u,v)=(\sigma,\sigma)$ of the elliptic coordinates, reaches another turning point $P_2$ with the same maximum value $u_M$ of the $u$-coordinate and finally asymptotically goes back to the origin $(u,v)=(\sigma,0)$. In sum, the orbit of the composite $Q$-balls transverses four times the interval $u\in [\sigma,u_M]$ and twice the interval $v=[0,\sigma]$. On the other hand, the $\widehat{\cal B}_1(x;\omega)$-orbit lies on the $u$-axis taking the values between $\sigma$ and $u_M$ twice. The $\overline{\cal B}_1(x;\omega)$-orbit transverses the segment $v\in [0,\sigma]$ with $u=\sigma$, crosses through the focus $F=(\sigma,\sigma)$, follows the segment $u\in [\sigma,u_M]$ with $v=\sigma$ fixed and subsequently goes back to the origin in the reverse direction. 

Before computing the total energy of the composite $Q$-balls, we shall prove the second of the statements, which says that the sum of the Noether charges $Q_1+Q_2$ is the same for all the members of the ${\cal B}_2(x;\omega,\gamma_1)$-family. The definition (\ref{charges}) of the $Q_i$-charges allows us to write in the $(u,v)$-plane
\begin{eqnarray*}
	&& \hspace{-0.5cm} Q_1 + Q_2 =  \omega \int_{-\infty}^\infty dx \Big[ u(x)^2 + v(x)^2 - \sigma^2 \Big]  =  \omega \int_{-\infty}^\infty dx \Big[  \frac{ u^2 (u^2 -\sigma^2)}{u^2 - v^2}  + \frac{v^2 (\sigma^2-v^2)}{u^2 - v^2} \Big] = \\
	&& = \omega \sum_{k=0}^N \int_{x_k}^{x_{k+1}} \Big| \frac{u}{\sqrt{u^4-a^2 u^2+ b^2-\omega^2}} \frac{du}{dx} \Big| dx  + \omega  \sum_{k=0}^N \int_{x_k}^{x_{k+1}} \Big| \frac{v}{\sqrt{v^4-a^2 v^2+ b^2-\omega^2}} \frac{dv}{dx}\Big|  dx = \\
	&& = \sum_{k=0}^N \Big| {\cal W}_u[u(x_{k+1})] -  {\cal W}_u[u(x_k)] \Big| + \sum_{k=0}^N \Big| {\cal W}_v[v(x_{k+1})] -  {\cal W}_v[v(x_k)] \Big|
\end{eqnarray*}
where
\[
{\cal W}_z(z) = \frac{\omega}{2} \log \Big[  a^2- 2 z^2 -2 \sqrt{z^4 - a^2 z^2 +b^2-\omega^2}  \Big] \hspace{0.4cm} .
\]
The first order differential equations (\ref{eledo}) have been used to find the exact differential of the previous integrands that defines the function ${\cal W}_z(z)$. Therefore,
\begin{eqnarray}
&& \hspace{-1cm} Q_1[{\cal B}_2(x;\omega,\gamma_1)] + Q_2[{\cal B}_2(x;\omega,\gamma_1)]= 4 {\cal W}_u(u_M) -2  {\cal W}_u(\sigma)-2 {\cal W}_v(0) =  \nonumber \\
& & = 2 {\cal W}_u(u_M)-2 {\cal W}_v(0)  +2 {\cal W}_u(u_M) -2  {\cal W}_u(\sigma) = Q_1[\overline{\cal B}_1(x;\omega)]  +  Q_2[\widehat{\cal B}_1(x;\omega)] 
\hspace{0.4cm} . \label{qtuv}
\end{eqnarray}


\noindent which justifies the sum rule between the $Q_i$-charges. Finally, the total energy of the composite $Q$-balls can be obtained as
\begin{eqnarray*}
&& \hspace{-1cm} E[{\cal B}_2(x;\omega,\gamma_1)]=  \overline{E}[{\cal B}_2(x;\omega,\gamma_1)] + \omega ( Q_1[{\cal B}_2(x;\omega,\gamma_1)] + Q_2[{\cal B}_2(x;\omega,\gamma_1)] ) = \\
&& =  4 W_u(u_M) -2  W_u(\sigma)-2 W_v(0) + \omega ( Q_1[{\cal B}_2(x;\omega,\gamma_1)] + Q_2[{\cal B}_2(x;\omega,\gamma_1)] )=   \\
&& = 2 W_u(u_M)-2 W_v(0) + \omega \, Q_1[\overline{\cal B}_1(x;\omega)]    +2 W_u(u_M) -2 W_u(\sigma) + \omega \, Q_2[\widehat{\cal B}_1(x;\omega)] = \\
&& = E[\overline{\cal B}_1(x;\omega)]  + E[\widehat{\cal B}_1(x;\omega)] 
\end{eqnarray*}
where we have used (\ref{Bogo02}) and that $W_z(z)$ is an increasing function in the interval $z\in (0,u_M)$. Note that $\frac{d W_z(z)}{dz} = z \sqrt{z^4-a^2 z^2 + b^2-\omega^2} >0$ in the previously mentioned interval.

\section{Stability analysis of the $Q$-balls in the model}

In this Section the stability of the $Q$-balls identified in Section \ref{sectionSol} will be studied following the now-standard approach on this topic initially developed in \cite{Friedberg1976}. In this seminal paper Friedberg, Lee and Sirlin analyze the classical stability of the $Q$-balls with respect to field fluctuations that maintain the value of the Noether charge $Q$ constant. The model addressed in this paper involves two complex scalar fields, so the solution at every point can be perturbed in two different channels. Two different subsections will be included below to discuss separately the stability of the single and composite $Q$-balls. As previously mentioned we will thoroughly follow the procedure introduced in \cite{Friedberg1976}. Details will only be provided to show some differences with respect to the standard approach.

\subsection{Stability analysis of the single $Q$-balls}

\begin{figure}[h]
	\centerline{\includegraphics[height=3.5cm]{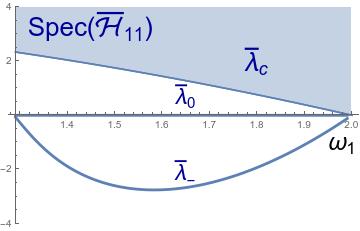} \hspace{0.5cm} \hspace{0.5cm}\includegraphics[height=3.5cm]{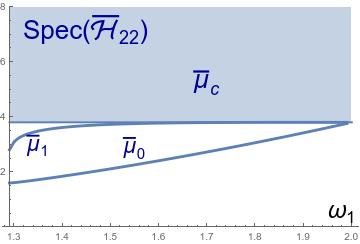} }
	\caption{\small Spectrum of the operators $\overline{\cal H}_{11}$ (left) and $\overline{\cal H}_{22}$ (right) for the parameter values $\sigma=0.25$, $a=1.75$, $b=2.0$ as a function of the internal rotation frequency $\omega_1$. } \label{fig:espectroQ1}
\end{figure}

We shall begin analyzing the stability of the single $\overline{\cal B}_1(x;\omega)$ type $Q$-balls. The standard approach consists in analyzing the behavior of the second variation of the energy functional $E[f_1,f_2]$ when small fluctuations $\delta f_1$ and $\delta f_2$ are respectively applied on each component of the solution $\overline{\cal B}_1(x;\omega)$. In this context, these perturbations must preserve the conserved Noether charges $Q_1$ and $Q_2$ defined in (\ref{charges}). This relates the $\delta f_1$-fluctuation to the variation $\delta \omega_1$ of the internal rotational frequency $\omega_1$ in the form
\[
\delta \omega_1 \int_{-\infty}^\infty f_1^2(x) dx = - 2\omega_1 \int_{-\infty}^\infty f_1 \delta f_1 dx \hspace{0.5cm} .
\]
Recall that the second component of the solution $\overline{\cal B}_1(x;\omega)$ vanishes and, consequently, $Q_2[\overline{\cal B}_1(x;\omega)]=0$. In order to preserve the value of $Q_2$ the fluctuations in the second component are assumed to be static. If we substitute the fluctuation $\delta F=(\delta f_1, \delta f_2)^t$ into the energy functional $E[f_1,f_2]$ the term which determines the behavior of the functional at second order is given by
\[
\delta E^{(2)}|_Q = \int_{-\infty}^\infty dx \frac{1}{2} (\delta F)^t \, {\cal H}[\overline{\cal B}_1(x)]\, \delta F + \frac{2\omega_1^3}{Q_1} \Big( \int_{-\infty}^\infty f_1 \delta f_1 dx \Big)^2
\]
where the second order small fluctuation operator ${\cal H}[\overline{\cal B}_1(x)]$ reads
\begin{equation}
{\cal H} [\overline{\cal B}_1(x)]= \left( \begin{array}{cc}
-\frac{d^2}{dx^2} + \left. \frac{\partial^2 U}{\partial f_1^2}  \right|_{\overline{\cal B}_1(x)} -\omega_1^2& \left.\frac{\partial^2 U}{\partial f_1 \partial f_2} \right|_{\overline{\cal B}_1(x)} \\[0.3cm]
\left.\frac{\partial^2 U}{\partial f_1 \partial f_2} \right|_{\overline{\cal B}_1(x)}  & -\frac{d^2}{dx^2} + \left.\frac{\partial^2 U}{\partial f_2^2} \right|_{\overline{\cal B}_1(x)}
\end{array}\right) \hspace{0.4cm}. \label{operator}
\end{equation}
For our model we have
\begin{eqnarray}
\frac{\partial^2 U}{\partial f_1^2} &=& b^2 -6a^2 f_1^2+15 f_1^4-2a^2f_2^2+18 f_1^2 f_2^2 + 3 f_2^4 + 2\sigma^2 f_2^2  \hspace{0.5cm} ,\label{der11pot} \\
\frac{\partial^2 U}{\partial f_1 \partial f_2} &=& - 4 a^2 f_1 f_2 + 12 f_1^3 f_2 +12 f_1 f_2^3 + 4 \sigma^2 f_1 f_2  \hspace{0.5cm} ,\label{der12pot} \\
\frac{\partial^2 U}{\partial f_2^2} &=&  b^2  -2a^2 f_1^2 + 3 f_1^4 - 6 a^2 f_2^2 + 18 f_1^2 f_2^2 + 15 f_2^4 - a^2 \sigma^2 + 2 \sigma^2 f_1^2 + 12 \sigma^2 f_2^2 +\sigma^4  \hspace{0.5cm} .\label{der22pot}
\end{eqnarray}
It can be checked that the expression (\ref{der12pot}) evaluated on the solution (\ref{qball01}) is equal to zero. This means that the fluctuation operator (\ref{operator}) is diagonal and the longitudinal and orthogonal perturbations $\delta f_1$ and $\delta f_2$ to the $\overline{\cal B}_1(x;\omega)$ solution are decoupled and evolve independently. The behavior of the longitudinal fluctuations are characterized by the eigenfunctions $\overline{\xi}_1^{(i)}$ of the spectral problem
\begin{equation}
\overline{\cal H}_{11} \, \overline{\xi}_1^{(i)} (z) = \frac{\overline{\lambda}_i}{4(b^2-\omega_1^2)} \overline{\xi}_1^{(i)} (z)  \label{spectralproblem1}
\end{equation}
where $\overline{\cal H}_{11} = \frac{1}{4(b^2-\omega_1^2)} \,{\cal H}_{11}[\overline{\cal B}_1(x,\omega_1)]$ is given by
\begin{equation}
\overline{\cal H}_{11} = -\frac{\partial^2 }{\partial z^2} + \frac{1}{4} - \frac{3 a^2}{a^2 + \sqrt{a^4-4(b^2-\omega_1^2)}\, \cosh z}  +  \frac{15 (b^2-\omega_1^2)}{(a^2 + \sqrt{a^4-4(b^2-\omega_1^2)}\, \cosh z)^2}  \label{operator1}
\end{equation}
with $z= 2\sqrt{b^2-\omega_1^2} \, x$. It can be checked that
\begin{equation}
\overline{\xi}_0(z) = \frac{\sinh z}{(a^2 + \sqrt{a^4-4(b^2-\omega_1^2)}\, \cosh z)^\frac{3}{2}} \label{zeromode1}
\end{equation}
is a zero mode of $\overline{\cal H}_{11}$ for any value of the internal rotation frequency $\omega_1$. The function (\ref{zeromode1}) has a node. This implies that there must be one and only one negative eigenvalue $\overline{\lambda}_-$. In Figure \ref{fig:espectroQ1}(left) the eigenvalues $\overline{\lambda}_i$ of the operator ${\cal H}_{11}[\overline{\cal B}_1(x,\omega)]$ (numerically identified for the representative model parameters $\sigma=0.25$, $a=1.75$, $b=2.0$) have been depicted as a function of the frequency $\omega_1$. Note the presence of only one negative eigenvalue $\overline{\lambda}_-$, which tends to zero as the rotational frequency $\omega_1$ approaches to the values $\omega_1^-$ and $\omega_1^+$. The continuous spectrum in this case emerges on the threshold value $b^2-\omega_1^2$.

At this point, the analysis of the stability of the $\overline{\cal B}_1(x;\omega)$ solutions versus $\delta f_1$-fluctuations is completely analogous to the study carried out by Friedberg, Lee and Sirlin in \cite{Friedberg1976}. Theorem 3 (stated there) can be directly applied to this class of perturbations. The necessary and sufficient conditions for $Q$-balls to be stable are: (1) the operator ${\cal H}_{11}[\overline{\cal B}_1(x;\omega)]$ must have only one negative eigenvalue and (2) the derivative of the Noether charge $Q_1$ with respect to the internal rotational frequency $\omega_1$ must comply with the relation $\frac{1}{Q_1} \frac{dQ_1}{d\omega_1} <0$. It has been shown that these two conditions are verified in this case. This implies that the $\overline{\cal B}_1(x)$-type $Q$-balls are stable with respect to longitudinal fluctuations.

On the other hand, the orthogonal eigenfluctuations $\overline{\xi}_2^{(i)}$ are governed by the  spectral problem
\[
\overline{\cal H}_{22} \, \overline{\xi}_2^{(i)} (z) = \frac{ \overline{\mu}_i}{4(b^2-\omega_1^2)} \overline{\xi}_2^{(i)}(z)
\]
where $\overline{\cal H}_{22} = \frac{1}{4(b^2-\omega_1^2)}  {\cal H}_{22}[\overline{\cal B}_1(x;\omega_1)]$ follows the expression
\begin{equation}
\overline{\cal H}_{22}  = -\frac{\partial^2 }{\partial z^2} + \frac{b^2-a^2\sigma^2+ \sigma^4}{4(b^2-\omega_1^2)} - \frac{a^2-\sigma^2}{a^2 + \sqrt{a^4-4(b^2-\omega_1^2)}\, \cosh z}  +  \frac{3 (b^2-\omega_1^2)}{(a^2 + \sqrt{a^4-4(b^2-\omega_1^2)}\, \cosh z)^2}  \label{h22long}
\end{equation}
Here, $z=2\sqrt{b^2-\omega_1^2}\, x$ and $\overline{\mu}_i$ are the corresponding eigenvalues of the original operator ${\cal H}_{22}[\overline{\cal B}_1(x;\omega_1)]$. They have been numerically calculated for the particular model parameters considered in this paper. In Figure \ref{fig:espectroQ1}(right) these eigenvalues $\overline{\mu}_i$ have been plotted as a function of the rotational frequency $\omega_1$. For the smallest values of $\omega_1$ there exist two discrete eigenvalues $\overline{\mu}_0$ and $\overline{\mu}_1$ whereas for large enough values of the frequency only one of them remains. The continuous spectrum in this case emerges on the threshold value $b^2-\sigma^2(a^2-\sigma^2)$. As we can see in Figure \ref{fig:espectroQ1}(right) there are no negative eigenvalues. In sum, the $\overline{\cal B}_1(x,\omega_1)$ solutions for the model parameters $\sigma=0.25$, $a=1.75$, $b=2.0$ are stable with respect to orthogonal perturbations. Given the form (\ref{h22long}) it is difficult to introduce a general analysis of the stability for any value of the parameters. Despite this fact, it will be proved below that the $\overline{\cal B}_1(x;\omega_1)$ balls are always stable when the asymmetry of the model is small enough. In order to do this, note that the operator
\begin{equation}
\widetilde{\cal H}_{22}  = -\frac{\partial^2 }{\partial z^2} + \frac{b^2-a^2\sigma^2+ \sigma^4}{4(b^2-\omega_1^2)} - \frac{a^2}{a^2 + \sqrt{a^4-4(b^2-\omega_1^2)}\, \cosh z}  +  \frac{3 (b^2-\omega_1^2)}{(a^2 + \sqrt{a^4-4(b^2-\omega_1^2)}\, \cosh z)^2}  \label{h22long2}
\end{equation}
has a ground state
\[
\widetilde{\xi}_0(x)= \frac{1}{(a^2 + \sqrt{a^4-4(b^2-\omega_1^2)}\, \cosh z)^\frac{1}{2}}
\]
with eigenvalue
\begin{equation}
\widetilde{\mu}_0 = \frac{\omega_1^2 - \sigma^2(a^2-\sigma^2)}{4(b^2-\omega_1^2)} \label{autocomp}
\end{equation}
which clearly is positive if $\sigma$ is small enough. It can be verified that
\[
\overline{\cal H}_{22} = \widetilde{\cal H}_{22} + \frac{\sigma^2}{a^2 + \sqrt{a^4-4(b^2-\omega_1^2)}\, \cosh z}
\]
which means that the potential well of $\overline{\cal H}_{22}$ is weaker than that in $\widetilde{\cal H}_{22}$. As a consequence, the eigenvalue of the ground state for the operator $\overline{\cal H}_{22}$ must be higher than $\widetilde{\mu}_0$ in (\ref{autocomp}). Therefore, the $\overline{\cal B}_1(x)$-type $Q$-balls are also stable with respect to orthogonal fluctuations for small values of $\sigma$.

Using a completely analogous reasoning as before and the continuity of the eigenvalues with respect to the model parameters it can be proved that the $\widehat{\cal B}_1(x)$ solutions (the second type of single $Q$-balls described in Section 3) are stable for small values of $\sigma$. Indeed, numerical analysis applied to the case $\sigma=0.25$, $a=1.75$ and $b=2.0$ leads to similar results to those found in Figure \ref{fig:espectroQ1}.

\subsection{Stability analysis of the composite $Q$-balls}

\begin{figure}[h]
	\centerline{\includegraphics[height=3.5cm]{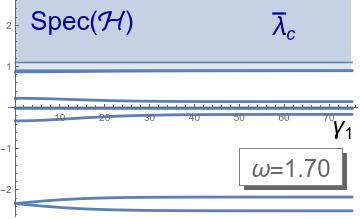} }
	\caption{\small Spectrum of the operator ${\cal H} [{\cal B}_2(x)]$ for the parameter values $\sigma=0.25$, $a=1.75$, $b=2.0$ and $\omega=1.70$ as a function of the family parameter $\gamma_1$. } \label{fig:espectroQcomp}
\end{figure}

The stability analysis is much more complicated for the composite $Q$-balls, where the two complex components of the solutions are non null. As before a deformation of the solution ${\cal B}_1(x;\omega,\gamma_1) + (\delta f_1,\delta f_2)$ which maintains the $Q_i$-charges constant is introduced into the energy functional $E[f_1,f_2]$. In this case, the variations of the two internal frequencies $\delta \omega_1$ and $\delta \omega_2$ must comply with the constraints
\[
\delta \omega_1 = -\frac{2\omega_1^2}{Q_1} \int_{-\infty}^\infty f_1 \delta f_1 dx \hspace{0.5cm},\hspace{0.5cm} \delta \omega_2 = -\frac{2\omega_2^2}{Q_2} \int_{-\infty}^\infty f_2 \delta f_2 dx \hspace{0.5cm} .
\]
If we substitute these fluctuations into the energy functional $E[f_1,f_2]$, the term at second order is given by
\begin{eqnarray}
&& \hspace{-0.8cm}\delta E^{(2)}|_Q =  \int_{-\infty}^\infty dx \frac{1}{2} (\delta F)^t \, {\cal H}[{\cal B}_2(x)]\, \delta F + \frac{2\omega_1^3}{Q_1} \Big( \int_{-\infty}^\infty f_1 \delta f_1 dx \Big)^2 + \frac{2\omega_2^3}{Q_2} \Big( \int_{-\infty}^\infty f_2 \delta f_2 dx \Big)^2 = \nonumber \\ 
&& = \int_{-\infty}^\infty dx \frac{1}{2} (\delta F)^t \, {\cal H}[{\cal B}_2(x)]\, \delta F + 2\omega^3 \Big( \int_{-\infty}^\infty (\delta F)^t \cdot \widetilde{F} dx \Big)^2 - 2 \int_{-\infty}^\infty dx \widetilde{F}_1 \delta f_1 \cdot \int_{-\infty}^\infty dx \widetilde{F}_2 \delta f_2 \label{hesscomp}
\end{eqnarray}
where $\delta F = (\delta f_1,\delta f_2)^t$ and $\widetilde{F} = (\frac{f_1}{\sqrt{Q_1}} , \frac{f_2}{\sqrt{Q_2}})^t$. The small fluctuation operator ${\cal H}[{\cal B}_2(x)]$ in (\ref{hesscomp}) reads
\begin{equation}
{\cal H} [{\cal B}_2(x)]= \left( \begin{array}{cc}
-\frac{d^2}{dx^2} + \left. \frac{\partial^2 U}{\partial f_1^2}  \right|_{{\cal B}_2(x)} -\omega_1^2& \left.\frac{\partial^2 U}{\partial f_1 \partial f_2} \right|_{{\cal B}_2(x)} \\[0.3cm]
\left.\frac{\partial^2 U}{\partial f_1 \partial f_2} \right|_{{\cal B}_2(x)}  & -\frac{d^2}{dx^2} + \left.\frac{\partial^2 U}{\partial f_2^2} \right|_{{\cal B}_2(x)}-\omega_2^2
\end{array}\right)   \hspace{0.5cm} . \label{operator2}
\end{equation}
It can be proved that now the existence of two negative eigenvalues in the spectrum of the operator (\ref{operator2}) is not a sufficient condition leading to the instability of the composite $Q$-balls because of the last term in (\ref{hesscomp}). The argument becomes valid again if three of these eigenstates are considered. Obviously, it is not possible to analytically solve the spectrum of the matrix operator (\ref{operator2}). Despite this fact it can be proved that this operator has two zero modes. Indeed, these eigenfunctions correspond to the expressions $\frac{\partial {\cal B}_2(x;\omega,\gamma_1)}{\partial x}$ and $\frac{\partial {\cal B}_2(x;\omega,\gamma_1)}{\partial \gamma_1}$. However, as far as we know, there are no mathematical results relating the nodes of these zero modes to the number of negative eigenvalues for matrix operators of the form (\ref{operator2}). Therefore, numerical analysis must be necessarily applied to obtain this information. In Figure \ref{fig:espectroQcomp} the spectrum of the operator ${\cal H} [{\cal B}_2(x)]$ as a function of the family parameter $\gamma_1$ has been depicted for the parameter values $\sigma=0.25$, $a=1.75$, $b=2.0$ and $\omega=1.70$. It can be observed that there are three negative eigenvalues, which from our previous claim implies that the composite $Q$-balls are unstable in this case although long-living. Similar numerical results have been found for other values of the parameters. The analysis of these instability channels could bring insight into the forces between the two single $Q$-balls rotating in each components of the internal space.

\section{Conclusions and further comments}

In this paper the existence of analytical solutions describing $Q$-balls in a family of deformed $O(4)$ sigma models in (1+1) dimensions has been investigated. These models involve two complex scalar fields whose coupling breaks the $O(4)$ symmetry group to $U(1)\times U(1)$. This leads to the existence of two Noether conserved $Q_i$-charges. The model parameter $\sigma$ can be understood as a measure of the deformation of the model with respect to the rotationally invariant theory. It has been shown that there are two types of single $Q$-balls rotating around each of the components of the internal space and a one-parameter family of composite $Q$-balls. All of these non-topological solitons have been analytically identified. The composite solutions consist of two single $Q$-balls (separated by a distance determined by the family parameter) spinning around each complex field with the same internal rotation frequency. Indeed, these solutions are formed by two energy lumps. It has been checked that the single $Q$-balls are linearly stable with respect to small fluctuations which preserve the $Q_i$-charges. The study of the stability for the composite solutions is more complicated. Following the arguments introduced in \cite{Friedberg1976} it has been proved that in this context the existence of three negative eigenvalues in the spectrum of the second order small fluctuation operator is a sufficient condition for proving the instability of these composite $Q$-balls. Numerical analysis has been used to analyze the evolution of these solutions when they are not initially perturbed. The results indicate that these composite solutions are long-lived. However, the numerical study of the Hessian operator (\ref{operator2}) shows the presence of three negative eigenvalues, which means that these solutions are unstable states. In this context it has been found that some perturbations make the two constituents of the $Q$-balls travel away or approach each other while the synchronization of the two internal rotation frequencies is lost. These results make evident the presence of forces between the different constituents of the solutions, which makes the non-topological solutions depend non-trivially on time.

The research introduced in the present work opens up some possibilities for future work. The results point out that the interaction between the single $Q$-balls described in Section 3 are highly non-trivial. For this reason it is very interesting to tackle the study of the scattering between these two types of solutions following, for example, the scheme introduced in \cite{Bowcock2009}. This could allow us to understand the forces between these constituents and the dependence of these forces with respect to the difference between the internal rotation frequencies of these $Q$-balls. The collision between excited $Q$-balls is also a future goal in order to discern if structure similar to those found in \cite{Alonso2021b, Campos2021-2} arise in this context.

\section*{Acknowledgments}

This research was funded by the Spanish Ministerio de Ciencia e Innovación (MCIN) with funding from the European Union NextGenerationEU (PRTRC17.I1) and the Consejería de Educación, Junta de Castilla y Le\'on, through QCAYLE project, as well as MCIN project PID2020-113406GB-I00 MTM.

\end{document}